\documentclass[12pt,onecolumn,peerreview]{IEEEtran}
\usepackage{fancyhdr}
\usepackage{epsfig}
\usepackage{threeparttable}
\usepackage{epsf,epsfig}
\usepackage{amsthm}
\usepackage{amsmath}
\usepackage{amssymb}
\usepackage{amsfonts}
\usepackage[noadjust]{cite}
\usepackage{dsfont}
\usepackage{subfigure}
\usepackage{color}
\usepackage{soul}
\usepackage{algorithmic,algorithm}

\newtheorem{lemma}{Lemma}

\newtheorem{proposition}{Proposition}

\def\qed{$\Box$}

\def\proof{\noindent{\emph{Proof:} }}

\def
\endproof{\hspace*{\fill}~\qed
\par
\endtrivlist\unskip}
\def\endproof{\hspace*{\fill}~\qed\par\endtrivlist\vskip3pt}

\def\E{\mathsf{E}}

\def\phi{\varphi}

\def\l{\left}
\def\r{\right}
\def\({\left(}
\def\){\right)}

\newcommand{\eref}[1]{(\ref{#1})}

\def\bff{{\mathbf{f}}}

\def\bv{{\mathbf{v}}}

\def\bx{{\mathbf{x}}}
\def\by{{\mathbf{y}}}

\def\b0{{\mathbf{0}}}

\def\bY{{\mathbf{Y}}}

\newcommand{\diag}{\mathrm{diag}}

\newcommand{\nn}{\nonumber}

\newtheorem {Remark}{Remark}

\begin{document}	
\title{Hybrid Beamforming via the Kronecker Decomposition for the Millimeter-Wave Massive MIMO Systems}
\author{Guangxu Zhu, Kaibin Huang, Vincent K. N. Lau, Bin Xia, Xiaofan Li and Sha Zhang
    \thanks{\setlength{\baselineskip}{13pt} \noindent G. Zhu and K. Huang are with the Dept. of EEE at The  University of  Hong Kong, Hong Kong (Email: huangkb@eee.hku.hk). V. K. N. Lau is with the Dept. of ECE at The Hong Kong University of Science and Technology, Hong Kong (Email: eeknlau@ece.ust.hk). Bin Xia is with the Dept. of EE at Shanghai Jiao Tong University, Shanghai, China (Email: bxia@sjtu.edu.cn). Xiaofan Li and Sha Zhang are with Shenzhen Institute of Radio Testing \& Tech., Shenzhen, China. (Email: lixiaofan@srtc.org.cn).}  
%\thanks{The research was supported by the Hong Kong Innovation and Technology Commission under the grant GHP01213SZ, the Shenzhen-Hong Kong Innovative Technology Cooperation Funding (Grant No. SGLH20131009154139588) and the Hong Kong Research Grants Council under the grant 17259416.}
}
\maketitle

\begin{abstract}
Millimeter-Wave (mmWave) massive multiple-input multiple-output (MIMO) seamlessly integrates two wireless  technologies, \emph{mmWave communications} and \emph{massive MIMO}, which provides spectrums with tens of GHz of total bandwidth and supports  aggressive space division multiple access using large-scale arrays, respectively.  Though it is  a promising solution for next-generation systems, the realization of mmWave massive MIMO faces several practical   challenges. In particular, implementing massive MIMO in the digital domain requires hundreds to thousands of radio frequency (RF) chains and analog-to-digital converters matching the number of antennas.  Furthermore, designing these components to operate at the mmWave frequencies is challenging and costly. These motivated the recent development of the \emph{hybrid-beamforming} architecture where MIMO signal processing is divided for separate implementation in the analog and digital domains, called the \emph{analog} and \emph{digital beamforming}, respectively. Analog beamforming using a phase array introduces uni-modulus constraints on the beamforming coefficients. They render the conventional MIMO techniques unsuitable and call for new designs. In this paper, we present a systematic design framework for hybrid beamforming for multi-cell multiuser  massive MIMO systems  over mmWave channels characterized by sparse propagation paths. The framework relies on the decomposition of analog beamforming vectors and path observation vectors into Kronecker products of factors being uni-modulus vectors. Exploiting properties of Kronecker mixed products, different factors of the analog beamformer are designed for either nulling interference paths  or coherently  combining data paths. Furthermore, a channel estimation scheme is designed for enabling the proposed hybrid beamforming. The scheme estimates the angles-of-arrival (AoA) of data and interference paths by analog beam scanning and data-path gains by analog beam steering. The performance  of the channel estimation scheme is analyzed. In particular, the \emph{AoA spectrum} resulting from beam scanning, which displays the magnitude distribution of paths over the AoA  range, is derived in closed-form. It is shown that the inter-cell interference level diminishes \emph{inversely} with the array size, the square root of pilot sequence length  and  the spatial separation between paths, suggesting different ways  of tackling pilot contamination. 
\end{abstract}

\begin{IEEEkeywords}
MmWave communications,  massive MIMO, analog beamforming, channel estimation, interference cancellation.\end{IEEEkeywords}

\section{Introduction}
\emph{Millimeter-wave} (mmWave) communications and massive \emph{multiple-input multiple-output} (MIMO) play key roles in enabling  gigabit wireless access  in next generation  communications systems \cite{andrews2014will}. 
The two technologies can be seamlessly integrated, leading to an active research area called \emph{mmWave massive MIMO}. Besides dramatically expanding the available bandwidth, communication in the mmWave frequency bands reduces the antenna spacing to the scale of millimeter and thereby allows a large number of antennas to be packed in a small volume \cite{SHanCM2015,SunTAP2013}, facilitating the implementation of massive MIMO. Provision base stations (BSs) with large-scale arrays enables sharp beamforming for suppressing propagation loss and spatial multiplexing for supporting  massive numbers of simultaneous users, which are the main advantages of massive MIMO \cite{RusekLarssonMarz:ScaleUpMIMO:2012}. 

Beamforming and other MIMO techniques are traditionally implemented in the digital domain \cite{GesShaETAL:FromTheoPrac:Apr:03,Gesbert:MultiCellMIMOCooperativeNetworks:2010}. This requires the attachment of a radio frequency (RF) chain and an analog-to-digital converter (ADC) to each antenna for digitalizing its observation. The design approach targeting small-scale MIMO is no longer practical for massive MIMO as provisioning a BS with hundreds of  RF chains and ADCs results in overwhelming cost and power consumption \cite{SunMag2014}. The issue is exacerbated by implementation in the  mmWave frequency range. This motivated the development of an increasingly popular approach, called \emph{hybrid beamforming}, that implements part of MIMO beamforming in the analog domain using a phase array, called \emph{analog beamforming} \cite{HeathCM2014,roh2014millimeter}. 
The analog beamforming is designed solving the said problem by dramatic channel-dimension reduction by exploiting channel sparsity and clustering structure. The phase-array implementation of analog beamforming introduces uni-modulus constraints on beamforming coefficients, making the classic design techniques based on linear algebra not directly applicable. In this paper, we present a systematic approach for designing   analog beamforming for interference cancellation and signal enhancement  and the matching channel estimation algorithm based on the Kronecker-product decomposition.

\subsection{Prior Work}
Recently, hybrid  beamforming (or hybrid precoding) introduced in \cite{el2014spatially,alkhateeb2014channel} has attracted strong interests from both academia and industry due to its potential for bridging the theory and practice for mmWave massive MIMO. As the classic digital MIMO techniques (e.g., eigenmode  and zero-forcing beamforming) cannot be directly applied, active research has been conducted on finding new solutions satisfying the analog-hardware constraints. For single-user massive MIMO systems, a simple approach is proposed in \cite{venkateswaran2010analog, zhang2005variable} for equivalent implementation of digital MIMO techniques in the analog domain based on the idea of realizing a  scaled version of  an arbitrary complex vector as the addition of two phase-shift vectors.  The approach is extended to multiuser systems in  \cite{weiyuJSTSP2016}. Focusing on only equivalent implementation, the approach fails to explore the sparsity structure of mmWave channels for complexity reduction and the resultant analog implementation may not be hardware efficient as it doubles the required phase shifters. To address this issue, an alternative design approach for hybrid beamforming, which adapts to channel sparsity, is proposed in \cite{el2014spatially}. Particularly, the hybrid beamforming design aims at approximating the optimal digital design by using iterative \emph{orthogonal-matching-pursuit} algorithm to minimize their Euclidean distance. The approximation is enhanced in \cite{yu2016alternating} by further incorporating manifold optimization.
%applies an iterative algorithm based on \emph{orthogonal matching pursuit} to minimize the Euclidean distance between the hybrid precoder and the optimal digital design (for the sake of spectral efficiency maximization). The algorithm is enhanced in \cite{yu2016alternating} by incorporating  manifold optimization. 
However, the existence of approximation error is inevitable and as a result the resultant approximation design is not capable of nulling interference.  Other implementation aspects  of hybrid beamforming  have also been reported in the existing literature. In particular, the  \emph{beamspace channel representation} introduced in \cite{BradyTAP2013} is adopted in \cite{WangJSAC2009,alkhateeb2015limited,hur2013millimeter} to represent sparse mmWave massive MIMO channels and design  beam-scanning codebooks for limited feedback. In addition, practical issues such as finite-resolution phase-shifters and the effects of partial CSI have been considered in \cite{HeathJSTSP2016} and \cite{alkhateeb2013hybrid} respectively for designing hybrid precoders. Prior work has focused on hybrid beamforming/precoding design for point-to-point or multiuser channels without considering inter-cell interference. { Inter-cell interference in mmWave systems may be less severe due to signal blockage by urban objects, and existing measurement results also suggest that typical mmWave channels tends to be noise limited \cite{RapcellularmmWave2013}. However, as next-generation networks are expected to have ultra-dense small cells, the dramatic shortening of inter-cell distances will reduce the blockage density in channels, resulting in sparse or even line-of-sight (LoS) interference channels \cite{andrews2016modeling}. This motivates the current work.}
% In view of prior work, there lacks a systematic approach for analog beamformer design incorporating inter-cell interference suppression for mmWave channels, motivating the current work. 

Channel estimation for hybrid beamforming is an area less explored but an important one with new research challenges. The main challenge arises from the fact that the digital estimator behind the analog beamformer has access only to  an effective MIMO channel with reduced dimensionality.  This limitation, referred to as the  \emph{channel subspace sampling limitation} in \cite{hur2013millimeter}, is first tackled in \cite{alkhateeb2014channel} by designing channel-estimation algorithms based on compressed-sensing theory. Specifically, by exploiting the sparsity in mmWave channels, the algorithmic design is formulated as a sparse-reconstruction problem by treating the hybrid beamformer as a measurement matrix. By scanning the beamformer over a given codebook, sub-space channel observations are obtained at the receiver, based on which the classic compressed-sensing techniques  can be used to recover the complete sparse  channel matrix. The compressed-sensing based  channel-estimation algorithms have been customized for different types of mmWave   systems including single-user outdoor systems \cite{berraki2014application} and multiuser systems with equal-gain LoS links  \cite{alkhateeby2015compressed}, and  tested in experiments using large-scale arrays with more than $1000$ antennas \cite{ramasamy2012compressive}. Despite providing a promising connection to the powerful tool of compressed-sensing, the existing designs may lead to  high complexity and are effective only under restrictive channel assumptions, limiting their practicality. According to standard compressed-sensing theory, accurate estimation of sparse channels under the said channel subspace sampling limitation is feasible only if the measurement matrix satisfied the so called \emph{restricted isometry property}. However, due to the uni-modulus constraints imposed on the analog beamforming coefficients, finding a proper measurement matrix (hybrid beamforming matrix) having such a property is extremely challenging.
%making efficient estimation difficult.
%This requires an exhaustive search over a large codebook for designing a proper measurement matrix having such a property, making efficient estimation difficult. 
Moreover, existing hybrid beamforming codebook designs based on, alternatively, adaptive compressed-sensing such as the multi-resolution hierarchical codebooks proposed  in \cite{alkhateeb2014channel} relied on oversimplifying channel assumptions such as single-LoS links and may suffer from severe error propagation in the estimation process for the multi-path channels as remarked in \cite{alkhateeb2014channel}. The drawbacks motivate the design of practical channel estimation algorithms for hybrid beamforming accounting for interference channels in the current work.

\subsection{Contributions and Organization}

We consider a typical cell in a  multi-cell cellular uplink  network where a base station (BS) equipped with a large-scale \emph{linear} antenna array serves  multiple single-antenna users via  a mmWave frequency band in the presence of inter-cell interference. The BS receives from the users either data streams or pilot signals in the communication and channel estimation phases, respectively, both of which are considered in the paper. For hardware efficiency,  the hybrid-beamforming architecture with a limited number of RF chains and ADCs is used at the BS to implement different massive MIMO techniques including interference cancellation, data detection and channel estimation. The typical model for a mmWave channel is adopted that is  characterized by finite paths specified by different angles-of-arrival (AoA) \cite{andrews2016modeling}. {In the model, the interference and signal are assumed to arrive at the BS from two non-overlapping sets of paths,  which can be justified by the fact that, due to the channel sparsity and the in general different physical locations of the interfering users and the intended users, the probability that an interfering path and an intended user path have exactly the same AoA is equal to zero.  The same assumption is also adopted in \cite{YinGesbert:CoordinatedApproachLargeScaleMIMO:2013}.}

In this paper, a hybrid-beamforming framework for mmWave massive MIMO uplink communication is developed. The framework is based on the Kronecker-product representations of channels and beamforming vectors, which were   originally proposed in  \cite{zhu2015analog} for designing another type of system, namely the one  for simultaneous wireless information-and-power transfer.  The key challenge for developing the said framework lies in designing the analog-beamforming vector  under the mentioned  uni-modulus constraints on the vector elements. The main contributions of the paper are  summarized as follows. 

\begin{itemize}
\item Consider a single user system. Given a single data stream, only one  RF chain is needed in the hybrid beamforming architecture for detecting the data stream, which reduces its design to that of an analog beamformer. The design objective is to maximize the signal power under the constraint of inter-cell interference nulling. Assume that the AoAs of the intended signal and inter-cell interference are known. Let the observation of a particular signal/interference path be called a data/interference path-vector. To achieve the objective, a systematic method for designing the analog beamformer, called \emph{Kronecker analog beamforming}, is proposed based on decomposing the beamforming vector as well as the data and interference path-vectors into corresponding Kronecker products of phase-shift factors, which are also vectors with uni-modulus elements. By exploiting  the \emph{mixed-product property} of the Kronecker products between the decomposed beamforming vector and the path-vectors, the design principle is to allocate a set of factors of the beamforming vector each for nulling a single interference path and the remaining factors for enabling coherent combining of the data paths for enhancing the intended signal power. The design procedure enforces the uni-modulus constraints on beamforming coefficients and thereby yields the desired analog beamformer.

\item The design is extended to a multiuser system where the hybrid-beamforming architecture has multiple RF chains for multiuser data stream detection. The resultant hybrid-beamforming  design cascades a high-dimension analog and a low-dimension digital beamformer. By generalizing the Kronecker-decomposition method for the single-user system,  the analog-beamforming matrix is designed for inter-cell interference nulling and data signal enhancement. The digital counterpart follows the conventional digital minimum-mean-square-error (MMSE) beamforming design to support spatial multiplexing by decoupling data streams from different intended users.

\item Targeting the hybrid-beamforming architecture, a scheme for efficient channel estimation  is proposed to  overcome the channel subspace sampling limitation and tackle  the well-known problem  of pilot contamination. The scheme comprises two steps: the first is to scan the AoA range using an analog beam for estimating the AoA of significant  signal/interference paths, followed by path-gain estimation. 
The path-AoA estimation involves peak detection of a \emph{AoA spectrum} displaying the magnitudes of different incident paths. The AoA spectrum is derived in closed-form that shows the inter-cell interference (pilot contamination level) diminishes \emph{inversely} with the increasing  array size, the square root of pilot sequence length  or  the minimum spatial separation between data and interference paths, suggesting  different  ways of coping with  pilot contamination. This is aligned with the main asymptotic result in \cite{YinGesbert:CoordinatedApproachLargeScaleMIMO:2013} but stronger and more elaborate. 

\item Next, given the estimated AoA, the gain estimation for the corresponding path essentially extracts a  path gain from the observed training signal by steering a corresponding analog beam based on either coherent-combining (CC) the targeted data path or zero-forcing (ZF) the inter-path interference. The path-gain estimation error is proved to  diminish as the array size increases with the decay rate of $O(\frac{1}{N})$ where $N$ denotes the number of antennas at BS. 

\end{itemize}

The remainder of the paper is organized as follows.  Section II introduces the system model. 
Section III presents the problem formulation for the hybrid beamforming design and the channel estimation. The proposed framework of Kronecker analog beamforming and the supporting channel-estimation scheme are presented in Section IV and Section V, respectively. Numerical results are provided in Section VI, followed by potential extensions and concluding remarks in Section VII and VIII respectively.

\section{System Model}

\begin{figure}[tt]
\centering
\includegraphics[width=10cm]{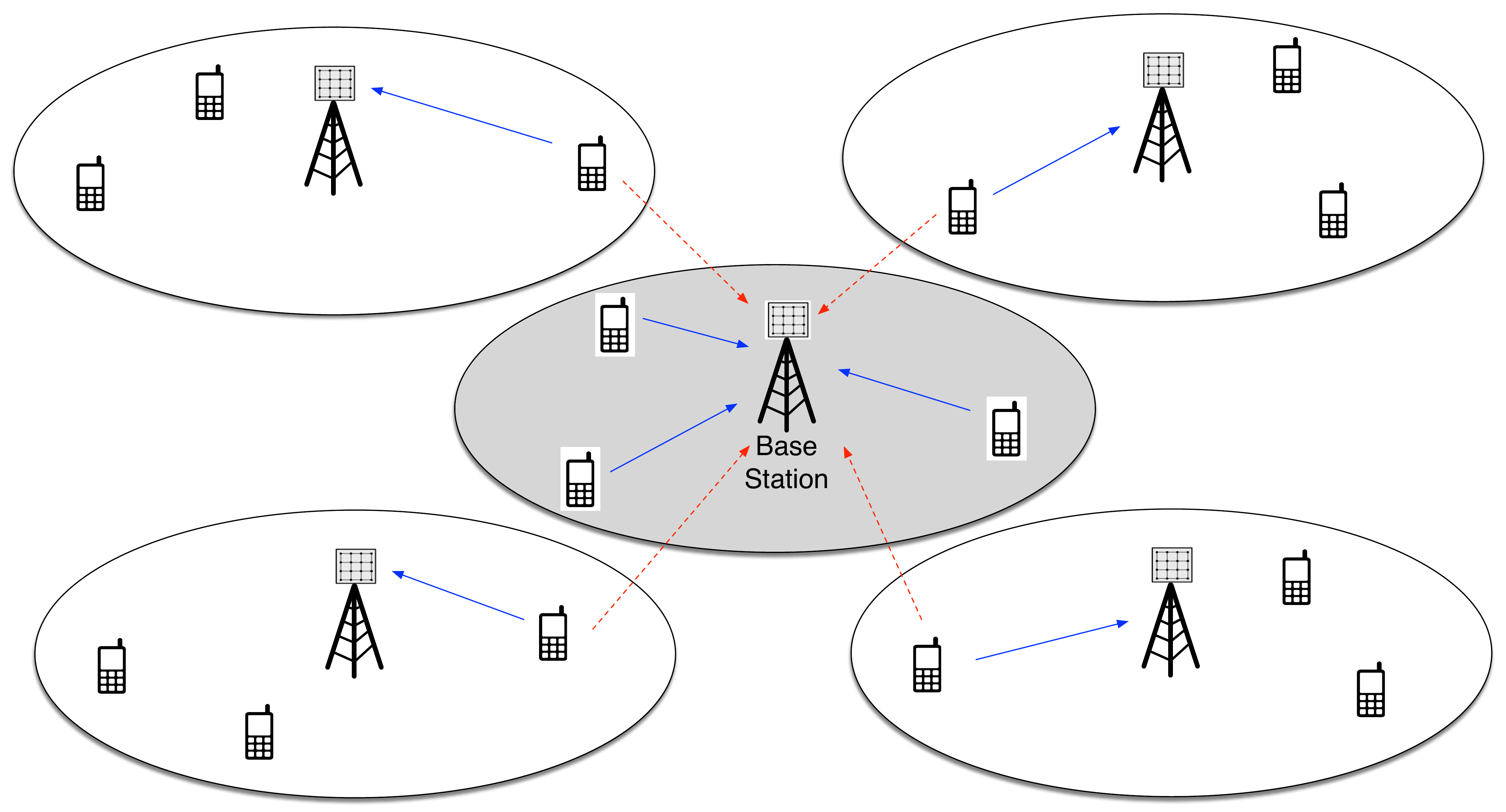}
\caption{A mmWave massive MIMO  system where uplink communication in the considered cell (shaded in gray) is exposed to inter-cell interference from the uplink users in nearby cells.}
\label{Fig:1}
\end{figure}

Consider a typical cell in a mmWave massive MIMO cellular system as shown in Fig. \ref{Fig:1}. A BS equipped with a large-scale uniform-linear antenna array of $N$ antennas communicates simultaneously with $K$ single-antenna users. We focus on the uplink but the proposed beamforming design can be straightforwardly extended  to the downlink by exploiting  the uplink-downlink duality. Due to the dense-BS deployment  and aggressive frequency reuse, the considered BS is exposed to inter-cell interference from $M$ uplink users in nearby cells. Hence, the received signal at the BS can be written as follows
\begin{align}\label{sm:1}
{\bf y} = {\bf G}{\bf x} + {\bf H}{\bf s} + {\bf n}.
\end{align}
The matrix ${\bf G} = [{\bf g}_1, {\bf g}_2, \cdots, {\bf g}_K]$ represents the data-channel  with ${\bf g}_k \in \mathbb{C}^{N \times 1}$ corresponding to the link between the $k$-th user and BS. Moreover, the interference-channel matrix is denoted as ${\bf H} = [{\bf h}_{1}, {\bf h}_{2}, \cdots, {\bf h}_{M}]$ where  ${\bf h}_{n} \in \mathbb{C}^{N \times 1}$ is the vector channel  between the $n$-th interferer and the BS. Let the symbol vectors transmitted by  the intended and interfering users be represented by ${\bf x} = [x_1,x_2,\cdots,x_K]^T$ and  ${\bf s} = [s_{1},  s_{2}, \cdots, s_{M}]^T$, respectively,  and their elements satisfy 
$\E[x_kx_k^*] = P_k$ and ${\sf E}[s_{n}s_{n}^*] = P'_{n}$ where $P_k$ and $P'_{n}$ are correspondingly the transmission power of the $k$-th intended user and the $n$-th interferer. The $N\times 1$ vector ${\bf n}$ is the \emph{additive white Gaussian noise} (AWGN) with ${\sf E}[{\bf nn}^H] = N_0{\bf I}$ where $N_0$ is the noise variance and ${\bf I}$ is an identity matrix. 

As mmWave signals can be blocked by objects  even of small sizes, the channels are characterized by sparse direct or reflected paths  \cite{andrews2016modeling}.  Therefore,  we adopt the multi-path model for both the data and interference channels. Specifically, the $k$-th data-channel vector comprises $L$ paths and is given as \cite{YinGesbert:CoordinatedApproachLargeScaleMIMO:2013}
\begin{align}\label{sys:1}
{\bf g}_k = \sum_{\ell=1}^L a_{k\ell}{\bf v}(\Phi_{k\ell}),
\end{align}  
where the coefficients for the $L$ paths,  $a_{k\ell}$, are  \emph{independent and identically distributed} (i.i.d.) ${\cal CN}(0,\delta_k^2)$ random variables with $\delta_k^2$ representing  the $k$-th data channel's propagation  attenuation. The  $N\times 1$ vector ${\bf v}(\Phi_{k\ell})$ represents the phase response for the  linear array at the BS:
\begin{align}\label{sys:2}
{{\bf v}(\Phi_{k\ell})} =[1,e^{j\Phi_{k\ell}},\cdots,e^{j(N-1)\Phi_{k\ell}}]^T,
\end{align}
where $\Phi_{k\ell} = \frac{2\pi d}{\lambda}\cos\phi_{k\ell}$ denotes the constant phase difference between the observations  by two adjacent antennas, $\phi_{k\ell}$ is the AoA, $d$ is  the antenna separation distance,  and $\lambda$ represents the carrier wavelength. {To simplify analysis, each interference-channel is assumed to contain a single dominant path with other paths omitted, which is justified by severe blockage for the interference-channel that has a relatively longer propagation distance than a data-channel.} To be specific,  
\begin{align}\label{sys:3}
{\bf h}_{n} = \beta_{n}{\bf v}(\Theta_{n}) = \beta_{n}[1,e^{j\Theta_{n}},\cdots,e^{j(N-1)\Theta_{n}}]^T,
\end{align} 
where  $\beta_n$ is  the channel coefficient, the function ${\bf v}$ is given in \eqref{sys:2}, and  $\Theta_{n}= \frac{2\pi d}{\lambda}\cos\theta_{n}$ is the inter-antenna phase difference with  $\theta_{n}$ being the AoA. 
{ Nevertheless, the proposed design framework is not limited to this assumption and can be easily extended to the general case that the interference-channel consists of multiple dominant paths.}

\begin{figure}[tt]
\centering
\includegraphics[width=12cm]{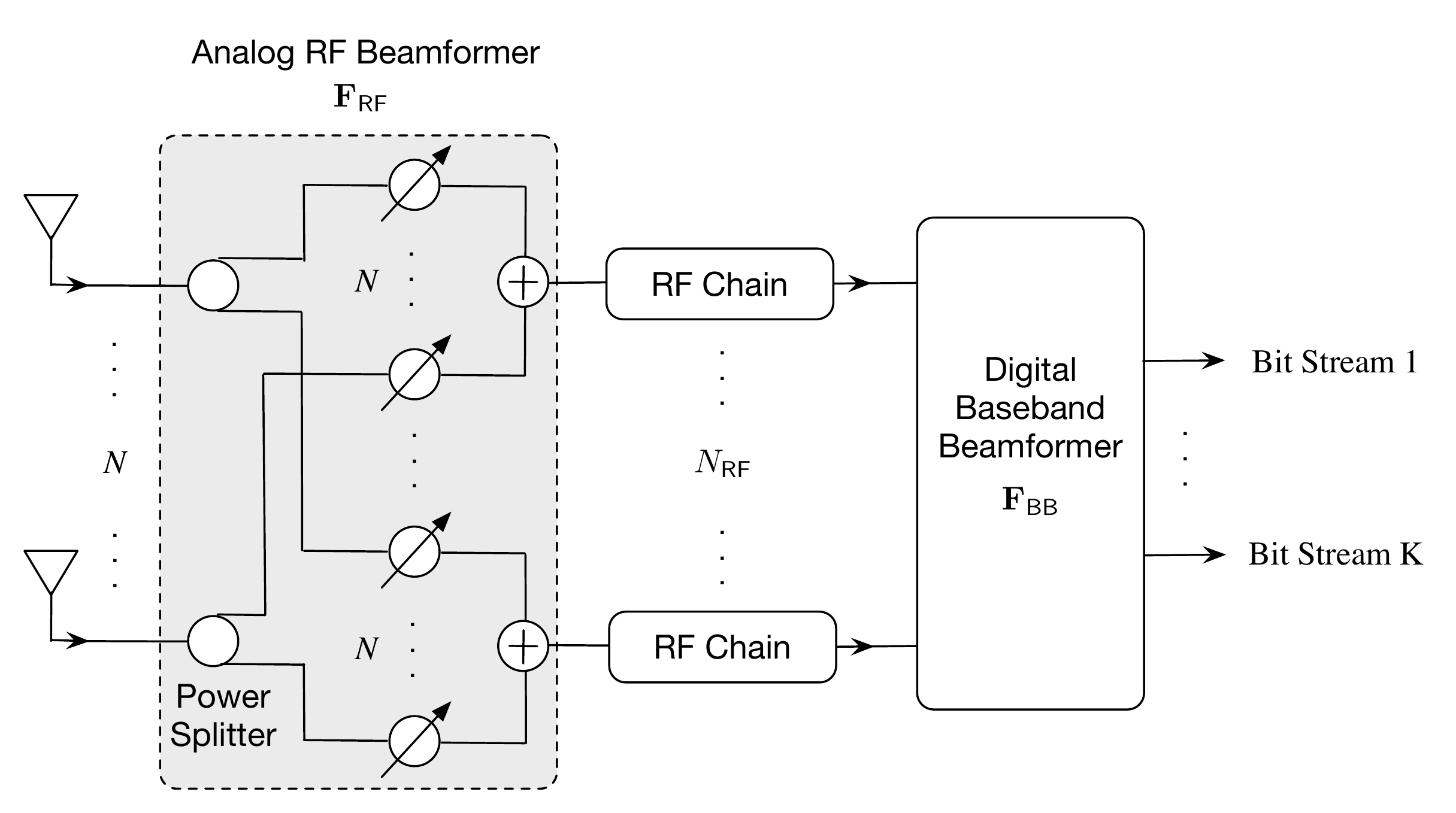}
\caption{ Hybrid beamforming architecture at the base station.}
\label{Fig:2}
\end{figure}

In the data-transmission phase, beamforming is performed at the BS for nulling inter-cell interference, enhancing SNRs for data streams and enabling spatial multiplexing. We consider the hybrid beamforming architecture  shown in Fig. \ref{Fig:2}, which cascades an $N\times N_{\sf RF}$  analog beamformer, denoted by ${\bf F}_{\sf RF}$, a band of $N_{\sf RF}$ RF chains,\footnote{Each RF chain comprises components such as low-noise amplifier, analog filter, downconverter  and ADC.}  and a $N_{\sf RF}\times K$ digital beamformer, denoted by ${\bf F}_{\sf BB}$. The analog beamformer is implemented using a phase  array of  continuously adjustable phase shifters which introduces uni-modulus constraints to the elements of ${\bf F}_{\sf RF}$.\footnote{It is worth pointing out that the additional RF circuit introduced by the implementation of analog beamforming may increase the chain noise figure at the receiver and thereby distort the signal quality to some extent \cite{Randomsignalandnoise}. To mitigate the noise figure elevation effect, low noise components should be used for analog beamforming implementation.} The function of the analog beamformer is to null inter-cell interference and enhance receive \emph{signal-to-noise ratios} (SNRs) for the data streams. The data streams will be further decoupled using the digital beamforming (without the uni-modulus constraints on its coefficients) which outputs $K$ parallel multiuser data streams for detection. Given the received signal in \eqref{sm:1}, the post-beamforming  signal can be written as 
\begin{align}
{\bar {\bf y}} = {\bf F}_{\sf BB}^H{\bf F}_{\sf RF}^H{\bf G}{\bf x} + {\bf F}_{\sf BB}^H{\bf F}_{\sf RF}^H{\bf H}{\bf s} + {\bf F}_{\sf BB}^H{\bf F}_{\sf RF}^H{\bf n}. 
\end{align}

In the channel-estimation phase,  users transmit pilot sequences with length of $Z$ symbols. Let the pilot sequence for the $k$-th user be denoted as ${\bf x}_k = [x_k(1), x_k(2), \cdots, x_k(Z)]^T$. Similarly, the $n$-th interfering pilot sequence from nearby cells is denoted by ${\bf s}_n = [s_n(1), s_n(2), \cdots, s_n(Z)]^T$. For tractability, several assumptions are made. First, all pilot sequences are binary sequences. {Second, the pilot sequences for the users in the considered cell are orthogonal and those received from the interfering users in nearby cells are approximated as randomly generated i.i.d. sequences to reflect the misalignment between the interfering pilots and the intended pilots due to the asynchornization between different cells. A similar assumption can also be found in \cite{Fernandes2013}.} As a result, the  pilot sequences used in different cells lack orthogonality, modelling pilot contamination. 
The  observation at the BS can be represented as a $N\times Z$ matrix ${\bf Y}$ given as
\begin{align}\label{aoa:2.1}
{\bf Y} = \sum_{k=1}^K{\bf g}_k{\bf x}_k^T + \sum_{n=1}^M{\bf h}_n{\bf s}_n^T +   {\bf N},
\end{align}
where ${\bf g}_k$ and ${\bf h}_n$ has been defined in (\ref{sys:1}) and (\ref{sys:3}) respectively, and ${\bf N} \in \mathbb{C}^{N \times Z}$ denotes the AWGN matrix. 

\section{Problem Formulation}\label{sec:III}

First, consider the joint design of analog beamformer with its digital counterpart. The objective is to maximize the sum-rate under the constraints of nulling inter-cell interference as well as the hardware limitation (i.e., the given number of RF chains and the phase-array implementation). Let the $N_{\sf RF} \times 1$ vector ${\bf f}_{{\sf BB}}(k)$ denote the $k$-th column vector of the digital beamformer ${\bf F}_{\sf BB}$. The achievable rate for user $k$ can be written as 
\begin{align}
R_k = \log_2\left({1 +  \frac{P_k|{\bf f}_{{\sf BB}}^H(k){\bf F}_{\sf RF}^H{\bf g}_k|^2}{\sum_{m \neq k}P_m |{\bf f}_{{\sf BB}}^H(k){\bf F}_{\sf RF}^H{\bf g}_m|^2 + \sum_{n=1}^MP'_{n}|{\bf f}_{{\sf BB}}^H(k){\bf F}_{\sf RF}^H{\bf h}_n|^2 + N_0\|{\bf f}_{{\sf BB}}^H(k){\bf F}_{\sf RF}^H\|^2}  }\right). 
\end{align}
The sum-rate is  then given by $R_{\sf sum} = \sum_{k=1}^KR_k$.  It follows that the sum-rate maximization problem can be formulated as follows
\begin{equation}({\bf P1})\qquad 
\begin{aligned}
\mathop {\max }\limits_{{\bf F}_{\sf RF}, {\bf F}_{\sf BB}} \; &{R_{\sf sum}} \\
{\textmd{s.t.}}\;\;&|[{\bf F}_{\sf RF}]_{p,q}| = 1,\;\forall p,q,\\ 
& \|{\bf F}_{\sf RF}^H{\bf h}_n\|^2 = 0, \ \forall n,
\end{aligned}
%\end{aligned}\nn
\label{P:1}
\end{equation}
where the first (uni-modulus) constraint arises from analog beamforming and the second one corresponds to interference nulling\footnote{Nulling the inter-cell interference in the analog domain benefits the quantization section by reducing the quantization noise and also alleviates the degrees of freedom requirement at the digital beamformer, thereby reduces the number of the required RF chains.}. The uni-modulus constraints make  the optimization problem non-convex and challenging to solve. Mathematically, due to the  constraints, the feasible set of P1, denoted as ${\mathbb S}$, is no longer the conventional  Euclidean vector space but a Riemannian manifold\footnote{Roughly speaking, Riemannian manifold is a kind of non-linear subspace with smooth surface (e.g, a spherical surface), in which the vectors no longer satisfy the closure properties of addition and scalar multiplication as those in Euclidean space \cite{RiemannianGeometry2008}.}, rendering the conventional optimization approach based on linear algebra  inapplicable here. A solution framework based on the proposed Kronecker decomposition technique will be presented in the following section for both the single-user and multi-user systems. 
Instead of the full knowledge of the entire channel matrices, i.e., ${\bf G}$ and ${\bf H}$, the solution requires only partial CSI, namely both the channel coefficients and AoA of data channels, $\bigcup_{k=1}^K\{ a_{k\ell}, \Phi_{k\ell}\}_{\ell=1}^L$,  and only the AoA of the interference channels, $\{\Theta_{n}\}_{n=1}^M$.\footnote{The knowledge of the interference path gains $\{\beta_n\}_{n=1}^M$ are not required by the proposed hybrid beamforming solution as presented in the sequel.} The proposed beamforming solution is built on the assumption that the said partial CSI is perfectly known at the BS. 

Next, to acquire the required CSI, the channel-estimation problem is formulated targeting the hybrid-beamforming architecture. In the channel training phase, all intended users transmit their pilot sequences to the BS for estimating the multiuser uplink channels. The received training signal at the BS is corrupted by not only noise but also inter-cell interference. This results in the so called  pilot contamination problem,  which poses a main challenge for accurate channel estimation and has no trivial solution. 
Given the channel models in  \eqref{sys:1} and \eqref{sys:2}, the channel estimation process is equivalent to inferring the corresponding path parameters (the AoAs and the path gains) from the observation matrix, $\bY$ in \eqref{aoa:2.1}, in the training duration: 
\begin{align}
{\bf Y}  \implies \l(\bigcup_{k=1}^K\underbrace{\{ a_{k\ell}, \Phi_{k\ell}\}_{\ell=1}^L}_{\textrm{Data Paths for User $k$}} \r)\cup \underbrace{\{\Theta_n\}_{n=1}^M}_{\textrm{Interference Paths}}. \label{Eq:Prob:Est}
\end{align}
As implied by \eqref{Eq:Prob:Est}, the estimation process also involves user identification of the paths by grouping  them for different users and tagging the rest as interference paths.  The problem is solved in Section \ref{sec:VI} by designing  a channel-estimation scheme matching the hybrid beamforming architecture.

\section{Analog Beamforming Design via Kronecker Decomposition}\label{sec:IV}
In this section, we solve Problem P1 by developing a hybrid beamformer solution via Kronecker decomposition of the analog beamforming and channel vectors. In particular,  we focus on the resource-limited scenario of $N_{\sf RF} = K$, which is the minimum number of RF chains required to support $K$ simultaneous users. {First, the designs of analog beamformers for single-user and multi-user systems are shown to be similar in the proposed solution. Thereby, for ease of notation and exposition, the Kronecker analog beamforming framework is then presented targeting the single-user system. Finally, the extension to the multiuser case as well as the implementation issues are discussed.
%and finally a design example is provided.
 }

\subsection{Similarity between the Single-user and Multiuser Cases}
\subsubsection{Single-user Case}
For the single-user  case with  $N_{\sf RF} = K =1$, the user's signal is exposed to only inter-cell interference but no intra-cell interference. 
The analog beamformer in Fig. \ref{Fig:2} is   a $N\times 1$ vector, represented  by ${\bf f}_{\sf RF}$, and the corresponding baseband beamformer is just a scalar which has no impact on the combined \emph{signal-to-interference-plus-noise ratio} (SINR). As a result, the beamformer design problem in P1 reduces to the  analog-beamformer design problem and can be rewritten as: 

\begin{equation}({\bf P2})\qquad 
\begin{aligned}
\mathop {\max }\limits_{ {\bf f}_{\sf RF} } \; & \log_2 \left({1 + \frac{P|{\bf f}_{\sf RF}^H{\bf g}|^2}{\sum_{n=1}^MP'_n \|{\bf f}_{\sf RF}^H{\bf h}_n\|^2 + N_0\|{\bf f}_{\sf RF}^H\|^2} }\right) \\
{\textmd{s.t.}}\;\;& |{\bf f}_{\sf RF}^H{\bf h}_n|^2 = 0, \ \forall n, \\
&|[{\bf f}_{\sf RF}]_{m}| = 1,\;\forall m.
\end{aligned}
%\end{aligned}\nn
\label{P:2}
\end{equation}
Under the interference nulling constraint and using the monotonically increasing property of the logarithmic  function, Problem  P2 can be further simplified as the following  signal-strength maximization problem: 

\begin{equation}({\bf P3})\qquad 
\begin{aligned}
\mathop {\max }\limits_{ {\bf f}_{\sf RF} } \; & |{\bf f}_{\sf RF}^H{\bf g}|^2 \\
{\textmd{s.t.}}\;\; & |{\bf f}_{\sf RF}^H{\bf h}_n|^2 = 0, \ \forall n, \\ 
&|[{\bf f}_{\sf RF}]_{m}| = 1,\;\forall m.
\end{aligned}\label{P:3}
\end{equation}
{ Note that the objective function in P3 is a convex function with respect to ${\bf f}_{\sf RF}$. Maximizing a convex function plus the non-convex feasible set due to the uni-modulus constraints makes the problem non-convex and challenging to solve. }

\subsubsection{Multiuser Case} In this case, different users' uplink transmissions are exposed to both intra-cell and inter-cell interference. The complex interference as well as the mentioned hardware constraints make Problem P1 more difficult to solve. For tractability, based on the hybrid-beamforming architecture in Fig. \ref{Fig:2}, Problem P1 is divided into sequential problems of designing the analog and digital beamformers separately\footnote{Despite the sequential design approach yields sub-optimality as it relaxes the coupling between the analog and digital beamformers, it makes the complex design problem tractable. Furthermore, it enables an elegant solution framework with low implementation complexity by allowing a separate design of each column of the analog beamforming matrix.}.  First, the analog beamformer is designed to cancel the inter-cell interference and also maximize the signal powers. Then the low-dimension digital beamformer is designed to decouple multiuser data streams by suppressing the intra-cell interference. Since inter-cell interference has been cancelled by the analog beamformer, the remaining multiuser decoupling can be conducted by a $K\times K$ digital beamformer, enabling the minimum RF chains implementation with $N_{\sf RF} = K$. Mathematically, the  analog-beamforming design problem is written as 

\begin{equation}({\bf P4})\qquad 
\begin{aligned}
\mathop {\max }\limits_{ {\bf f}_{\sf RF}(k) } \; & |{\bf f}_{\sf RF}^H(k){\bf g}_k|^2, \;\; k = 1,2,\cdots,K,
\\ 
{\textmd{s.t.}}\;\; & |{\bf f}_{\sf RF}^H(k){\bf h}_n|^2 = {0},\;\forall n\\ 
&|[{\bf f}_{\sf RF}(k)]_{m}| = 1,\;\forall m,
\end{aligned}
%\end{aligned}\nn
\label{P:4}
\end{equation}
where ${\bf f}_{\sf RF}(k)$ denotes the $k$-th column of the analog-beamforming matrix ${\bf F}_{\sf RF}$. The problem aims at  maximizing the $k$-th user's signal strength under the inter-cell interference nulling and uni-modulus constraints. Comparing Problems P3 and P4, we note that the design of each column of ${\bf F}_{\sf RF}$ for the multiuser case and that of ${\bf f}_{\sf RF}$ for the single user case are observed to have the identical form.

Given the analog beamformer that solves Problem P4, denoted as ${\bf F}^*_{\sf RF} = [{\bf f}_{{\sf RF}}^*(1), {\bf f}_{{\sf RF}}^*(2),\cdots, {\bf f}_{\sf RF}^*(K)]$, its digital counterpart ${\bf F}_{\sf BB}$ can be designed to further enhance the SINR of each data stream (as expressed in Problem P1) as follows
\begin{equation}({\bf P5})\qquad 
\begin{aligned}
\mathop {\max }\limits_{ {\bf f}_{{\sf BB}}(k) } \; & \frac{P_k|{\bf f}_{{\sf BB}}^H(k)({\bf F}_{\sf RF}^*)^H{\bf g}_k|^2}{\sum_{m \neq k}P_m |{\bf f}_{{\sf BB}}^H(k)({\bf F}_{\sf RF}^*)^H{\bf g}_m|^2 + N_0\|{\bf f}_{\sf BB}^H(k)({\bf F}_{\sf RF}^*)^H\|^2}, \;\; k = 1,2,\cdots,K,
\end{aligned}
%\end{aligned}\nn
\label{P:5}
\end{equation}
where ${\bf f}_{{\sf BB}}(k)$ denotes the $k$-th column of the digital beamforming matrix  ${\bf F}_{\sf BB}$. This is a classical unconstrained SINR maximization problem with colored noise, which can be solved by the following MMSE beamforming\footnote{The generalization to the weighted sum rate maximization for ensuring fairness among users can be done by designing the digital beamformer using the weighted MMSE solution proposed in \cite{WeightedMMSE2011} instead, while the design of the analog beamformer remains unchanged.}: 
\begin{align}\label{mmse:1}
{\bf F}_{\sf BB} = (\tilde {\bf G}{\bf D}\tilde {\bf G}^H + N_0({\bf F}_{\sf RF}^*)^H{\bf F}_{\sf RF}^*)^{-1}\tilde {\bf G},
\end{align}
where $\tilde {\bf G} = ({\bf F}_{\sf RF}^*)^H {\bf G}$ denotes the effective data-channel matrix observed by the digital baseband after analog beamforming, ${\bf D}$ is a diagonal matrix with the diagonal elements collecting the transmit power of the intended users, i.e.,  $[{\bf D}]_{k,k} = P_k$. 

In summary, the proposed hybrid-beamforming design comprises the small-scale MMSE digital beamforming and the large-scale analog beamforming whose design is equivalent to that of the single-user case. This equivalence allows the subsequent design of the analog-beamforming framework to focus on the single-user case for notation simplification.

\subsection{The Framework of Kronecker Analog Beamforming}\label{sec:IV:B}
The proposed framework of Kronecker analog beamforming that solves Problem P3 comprises three key steps. First, the analog beamforming and interference/data path vectors are decomposed into Kronecker products of phase-shift vectors. Based on the decomposition, in the following two steps, the analog beamformer is designed to null inter-cell interference and then enhance data-signal strength. The design procedure is summarized in Algorithm~\ref{algorithm:2} with the three steps elaborated in the following.

\subsubsection{Step $1$: Kronecker Decomposition of Analog Beamforming and Path Vectors} 
The method of vector Kronecker decomposition is described as follows. Let $\bv$ be a $N\times 1$ vector with uni-modulus elements and having the Vandermonde structure: $\bv = [1, e^{j\Theta}, e^{j2\Theta}, \cdots, e^{j(N-1)\Theta}]^T$ with $\Theta$ fixed. 
\begin{lemma}[Kronecker decomposition \cite{zhu2015analog}]\label{prop:1}\emph{
Given $N = n_1n_2\cdots n_{D}$ with $\{n_m \}_{m=1}^D$ being positive integers, the vector  ${\bf v}$ can be decomposed as 
\begin{align}\label{eq:p1}
{\bf v} = {\bf v}^{(1)} \otimes  {\bf v}^{(2)} \otimes \cdots \otimes {\bf v}^{(D)},
\end{align}
where $\otimes$ represents the left Kronecker product \cite{HorJoh:MatrAnal:85} and the $m$-th factor having the length of $n_m$ is given by
\begin{equation}\nn
{\bf v}^{(m)} = \l[1,e^{jn_{m - 1}\cdots n_1n_0\Theta},e^{j2n_{m - 1}\cdots n_1n_0\Theta}\cdots,e^{j(n_m - 1)n_{m - 1}\cdots n_1n_0\Theta}\r],
\end{equation}
 with $n_0 = 1$.} 
 \end{lemma}

The key property of Kronecker decomposition is that the vector $\bv$ and its factors are all phase-shift vectors with uni-modulus elements. Using Lemma \ref{prop:1}, the interference path vector ${\bf v}(\Theta_n)$ in \eqref{sys:3} can be decomposed as 
\begin{equation}\label{Eq:Path:Kronecker}
{\bf v}(\Theta_n) = {\bf v}^{(1)}_n \otimes  {\bf v}^{(2)}_n \otimes \cdots \otimes {\bf v}^{(D)}_n,
\end{equation}
where the factors are called \emph{Kronecker factors}, 
\begin{equation}\nn
{\bf v}^{(m)}_n = \l[1,e^{jn_{m - 1}\cdots n_1n_0\Theta_n},e^{j2n_{m - 1}\cdots n_1n_0\Theta_n}\cdots,e^{j(n_m - 1)n_{m - 1}\cdots n_1n_0\Theta_n}\r].
\end{equation}
Furthermore, the  analog beamforming vector ${\bf f}_{\sf RF}$ in Problem P3 is designed as the Kronecker product of phase-shift factors as follows: 
 \begin{equation}
\textrm{(Kronecker Analog Beamformer)}\qquad {\bf f}_{\sf RF} =  {\bf f}^{(1)} \otimes {\bf f}^{(2)}\otimes \cdots \otimes{\bf f}^{(D)},\label{kd:2}
\end{equation}
where the $m$-th Kronecker factor is given by ${\bf f}^{(m)} = [e^{j\psi_{m,1}},e^{j\psi_{m,2}},\cdots,e^{j\psi_{m, n_m}}]$ with $\{\psi_{m, n}\}$ being beamformer variables. The Kronecker factors  $\{{\bf f}^{(m)}\}$ represent the \emph{degrees-of-freedom} (DoF) in the analog beamformer. The basic idea of this design is to exploit the \emph{mixed product property} of Kronecker product and adjust the beamformer variables for inter-cell interference cancellation and signal strength enhancement as elaborated in the sequel. 
As an example, for the special case of  $N = 2^D$, given the prime factorization of $N$ as $N = 2\times 2\times \cdots \times2$, the interference path vector ${\bf v}(\Theta_n)$ and analog beamformer ${\bf f}_{\sf RF}$ can be decomposed to $D$ Kronecker factors as: 
\begin{align}
{\bf v}(\Theta_n) &= \l[1,e^{j\Theta_n}\r] \otimes  \l[1,e^{j2\Theta_n}\r] \otimes \cdots \otimes \l[1,e^{j2^{D-1}\Theta_n}\r],\label{Eq:Path:Kro}\\
{\bf f}_{\sf RF} &=  [e^{j\psi_{1,1}},e^{j\psi_{1,2}}] \otimes [e^{j\psi_{2,1}},e^{j\psi_{2,2}}]\otimes \cdots \otimes[e^{j\psi_{D,1}},e^{j\psi_{D,2}}]. \label{Eq:Beam:Kro}
\end{align}
Last, it is worth mentioning that a similar  technique was adopted in \cite{zhu2015analog} to design an analog spatial canceller for a wirelessly powered communications system targeting a different purpose of decoupling the strong  simultaneous wireless information and power transfer (SWIPT) signals over LoS channels and the weak wireless communication signals over rich scattering channels. 
\begin{Remark}
\emph{ Although the solution approaches in the current work and \cite{zhu2015analog} both leverage the Kronecker representations of channels and beamforming vectors, they differ in two main aspects. First, they solve different problems. The current work aims at solving a sum-rate maximization problem, which involves designing techniques for interference nulling and achieving both diversity and multiplexing gains. In contrast, a rank-maximization problem is tackled in \cite{zhu2015analog} via the constructions of a set of linearly independent vector bases. Next, substantial difference between the detailed designs in the current work and \cite{zhu2015analog} arise from the consideration of different systems and corresponding channel models, namely that one is a multi-cell mmWave system and the other a system for wireless power transfer.}
\end{Remark} 

\begin{Remark} [Optimal Kronecker Decomposition]
\emph{As indicated in Lemma \ref{prop:1}, different factorizations of $N$ lead to different decompositions of the analog beamformer ${\bf f}_{\sf RF}$. Among all, the \emph{prime factorization} of $N$ (i.e., $\{n_m \}$ are prime numbers) yields the largest number of factors, maximizing the DoF in the  resultant analog beamformer. Thus the prime factorization of $N$ should be used in the design. }
\end{Remark} 

\begin{algorithm}[tt]
\begin{itemize}
\item[0)] Decompose the array size $N$ into a product of integers: $N = n_1 n_2\cdots n_D$. 
\item[1)] Perform Kronecker decomposition of  the interference path vectors using \eqref{Eq:Path:Kronecker} and represent the analog beamforming vector ${\bf f}_{\sf RF}$ in a similar form using \eqref{kd:2}. 
\item[2)] Design the first $M$ Kronecker factors of ${\bf f}_{\sf RF}$ for inter-cell interference nulling using \eqref{in:2}. 
\item[3)] Design the remaining  $(D-M)$ Kronecker factors of ${\bf f}_{\sf RF}$ for data-signal enhancement  using \eqref{ue:3}. 
\end{itemize}
\caption{Summary of the Design Procedure for Kronecker  Analog Beamforming}\label{algorithm:2}
\end{algorithm}

\subsubsection{Step $2$: Analog Interference Cancellation} 
Based on the Kronecker decomposition expressions in (\ref{Eq:Path:Kronecker}) and (\ref{kd:2}), the inter-cell interference nulling constraints in P3 can be rewritten using the \emph{mixed-product property} of Kronecker product as follows.
\begin{align}\label{in:01}
{\bf f}_{\sf RF}^H{\bf h}_n = 0 \;\; &\iff \;\;  {\bf f}_{\sf RF}^H{\bf v}(\Theta_n) = 0 \notag\\  
&\iff \;\; ({\bf f}^{(1)})^H{\bf v}_{n}^{(1)}\otimes ({\bf f}^{(2)})^H{\bf v}_{n}^{(2)}\otimes \cdots \otimes ({\bf f}^{(D)})^H{\bf v}_{n}^{(D)} = 0,\;\;\;\;\forall n\in\{1,2,\cdots,M\}.
\end{align} 
A key observation that motivates the proposed design is that  a particular interfering path ${\bf v}(\Theta_n)$ can be cancelled by setting any factor in  \eqref{in:01} to be zero, namely
\begin{equation}\label{in:02}
({\bf f}^{(m)})^H{\bf v}_{n}^{(m)} = 0.\;\;\;\;\exists m\in\{1,2,\cdots,D\}.
\end{equation} 
In other words, cancelling  interference from all $M$ paths requires $M$ out of total $D$ factors (or DoF) of the analog beamformer  ${\bf f}_{\sf RF}$.\footnote{ Note that due to the uni-modulus constraints on the coefficients of ${\bf f}^{(m)}$, and the prime number constraints on its length (see Remark 2), ${\bf f}^{(m)}$ can  be designed to be orthogonal to ${\bf v}_{n}^{(m)}$ for only one $n$ value, leading to the said requirement.} This requires that $D$ is larger than $M$. The requirement can be satisified by mmWave massive MIMO systems since  the number of antennas is typically  far larger than the number of interference paths. Without loss of generality, the first $M$ Kronecker factors of ${\bf f}_{\sf RF}$ are assigned for inter-cell interference cancellation with the $m$-th factor targeting  the $m$-th interfering path ${\bf v}(\Theta_m)$. It follows that  the constraints in \eqref{in:02} can be rewritten as
\begin{equation}\label{in:1}
({\bf f}^{(m)})^H{\bf v}_{m}^{(m)} = 0,\;\;\;\; m\in\{1,2,\cdots,M\}.
\end{equation} 

\begin{Remark}[Selection of Kronecker Factors for Interference Nulling]
\emph{It can be observed from \eqref{in:02} that choosing arbitrary $M$ beamformer Kronecker factors is sufficient for nulling $M$ interference paths regardless of the vector lengths of the chosen factors. Thus it is desirable to choose $M$ shortest factors for this purpose so as to maximize the number of remaining beamformer variables for signal enhancement in Step~$3$ in the sequel. Without loss of generality, assume the factors of $N$ are monotone increasing: $n_1 \leq n_2 \cdots \leq n_{D}$. Then the first $M$ factors having the shortest lengths are chosen for inter-cell interference nulling and the remaining $(D-M)$ factors are for data-signal enhancement.}
\end{Remark}

There exist various methods for designing the phase-shift beamformer factors ${\bf f}^{(m)}$ to satisfy the interference-nulling constraints in \eqref{in:1} with one discussed  as follows. By exploiting the row-orthogonality and uni-modulus elements of a Fourier matrix, each of the beamformer factor can be designed to have the following form: 
\begin{align}\label{in:2}
 {\bf f}^{(m)} = \diag\l({\bf v}_{m}^{(m)}\r){\bf t}_{m},\;\;\; m\in\{1,2,\cdots,M\}, 
\end{align}
where $\diag\l({\bf v}_{m}^{(m)}\r)$ is a $n_m\times n_m$ diagonal matrix whose diagonal  elements are taken from those of ${\bf v}_{m}^{(m)}$, and ${\bf t}_{m}^T$ is a row vector arbitrarily selected from the  rows of the $n_m\times n_m$ Fourier matrix except the first one (an all-one vector). This design satisfies  the constraints in \eqref{in:1} as shown below: 
\begin{equation}
({\bf f}^{(m)})^H{\bf v}_{m}^{(m)} = {\bf t}_{m}^H \mathbf{1}=0, \qquad   m\in\{1,2,\cdots, M\},
\end{equation} 
where $\mathbf{1}$ represents the all-one vector with length of $n_m$. Note that the first equality arises from ``match filtering" the interference path  ${\bf v}_{m}^{(m)}$ and the second one is due to the row orthogonality of the Fourier matrix.  For the case that $n_m$ is equal to $2$ or a multiple of $4$, ${\bf t}_{m}^T$ can be also selected from the rows of a $n_m \times n_m$ Hadamard matrix which has only binary elements of $1$ and $-1$, simplifying practical implementation.

\subsubsection{Step $3$: Analog Signal Enhancement} Given the first $M$ Kronecker factors of the analog beamformer have been designed for interference cancellation in Step~$2$, in the current step, the remaining $(D-M)$ factors,  ${\bf f}^{(M+1)}, \cdots, {\bf f}^{(D)}$, are designed  for maximizing the data-signal strength as follows. Instead of designing these factors one by one, it is more convenient to directly design their Kronecker product, denoted by ${\bf f}_{\sf eq}^{(M+1)}$ and given as 
\begin{align}
{\bf f}_{\sf eq}^{(M+1)} = {\bf f}^{(M+1)} \otimes {\bf f}^{(M+2)} \otimes \cdots \otimes {\bf f}^{(D)}.
\end{align}
As a result, the desired analog beamforming vector can be rewritten as
\begin{align}\label{SE_factor}
{\bf f}_{\sf RF} =  {\bf f}^{(1)} \otimes {\bf f}^{(2)}\otimes \cdots \otimes{\bf f}^{(M)} \otimes {\bf f}_{\sf eq}^{(M+1)}. 
\end{align}
To facilitate the design, the data path vectors $\{{\bf v}(\Phi_{\ell})\}$ are decomposed in the same form as in \eqref{SE_factor}: 
\begin{align}\label{ue:0}
{\bf v}(\Phi_{\ell}) = {\bf u}_{\ell}^{(1)} \otimes  {\bf u}_{\ell}^{(2)} \otimes \cdots \otimes {\bf u}_{\ell}^{(M+1)}, \qquad \ell = 1, 2, \cdots, L
\end{align}
where the Kronecker factors match those  in \eqref{SE_factor} in terms of vector lengths. By substituting \eqref{SE_factor} and \eqref{ue:0} into \eqref{P:3} and applying the mixed-product property of Kronecker product, the objective function of Problem P3 is rewritten as 
\begin{align}\label{ue:1}
\l|{\bf f}_{\sf RF}^H{\bf g}\r|^2 = \left|{\sum_{\ell=1}^L a_\ell  ({\bf f}_{\sf eq}^{(M+1)})^H {\bf u}_{\ell}^{(M+1)} \prod_{m=1}^{M} ({\bf f}^{(m)})^H {\bf u}_{\ell}^{(m)}}\right|^2. 
\end{align}
Given the Kronecker factors  ${\bf f}^{(1)},\cdots,{\bf f}^{(M)}$ fixed, to simplify notation, define
\begin{equation}\label{ue:2}
\tilde a_\ell = a_\ell \prod_{m=1}^{M} ({\bf f}^{(m)})^H {\bf u}_{\ell}^{(m)}
\hspace{1em} \text {and} \hspace{1em} 
\tilde {\bf g} = \sum_{\ell=1}^L\tilde a_\ell {\bf u}_{\ell}^{(M+1)}
\end{equation} 
that respectively denote the $\ell$-th effective data-path gain and the effective  data-channel vector aggregating $L$ paths, conditioned on inter-cell interference nulling. Substituting (\ref{ue:2}) into (\ref{ue:1}), Problem P3 can be transformed into the following equivalent form.
\begin{equation}({\bf P6})\qquad 
\begin{aligned}
\mathop {\max }\limits_{ {\bf f}_{\sf eq}^{(M+1)} } \; & |({\bf f}_{\sf eq}^{(M+1)})^H{\tilde {\bf g}}|^2 \\
{\textmd{s.t.}}\;\; &|[{\bf f}_{\sf eq}^{(M+1)}]_{i}| = 1,\;\forall i.
\end{aligned}
%\end{aligned}\nn
\label{P:6}
\end{equation}
Since the elements of ${\bf f}_{\sf eq}^{(M+1)}$ are all uni-modulus, it can be written as
${\bf f}_{\sf eq}^{(M+1)} = [e^{j\Omega_1},e^{j\Omega_2},\cdots,e^{j\Omega_{n_{M+1}}}]^T$, where $n_{M+1}$ is the length of ${\bf f}_{\sf eq}^{(M+1)}$. Moreover, expand $\tilde {\bf g}$  as  $\tilde {\bf g} = [\tilde g_1, \tilde g_2,\cdots, \tilde g_{n_{M+1}}]$. By the  \emph{vector triangle inequality}, we have
\begin{align}
|({\bf f}_{\sf eq}^{(M+1)})^H{\tilde {\bf g}}|^2 = \left|{\sum_{i=1}^{n_{M+1}}e^{-j\Omega_i}\tilde g_i}\right|^2 \leq \left|{\sum_{i=1}^{n_{M+1}}|\tilde g_i|}\right|^2,
\end{align}
where the equality holds when $\Omega_i$ is equal to the phase of $\tilde g_i$. Therefore, the optimal solution of Problem P6 is 
\begin{align}\label{ue:3}
{\bf f}_{\sf eq}^{(M+1)} = [e^{j \angle {\tilde g_1}},e^{j \angle {\tilde g_2}},\cdots,e^{j \angle {\tilde g_{n_{M+1}}}}]^T.
\end{align}
Substituting \eqref{ue:3} into \eqref{SE_factor} completes the design of analog beamformer.

\subsection{Discussion}
Extension of the proposed design framework of Kronecker analog beamforming to the multiuser case and various implementation issues are discussed as follows. 

\subsubsection{Extension to the Multiuser Case} For the current case, the analog beamformer ${\bf F}_{\sf RF}$ is a matrix with uni-modulus elements, which  can be written in terms of its column vectors as 
\begin{equation}
{\bf F}_{\sf RF} = [{\bf f}_{{\sf RF}}(1), {\bf f}_{{\sf RF}}(2), \cdots ,{\bf f}_{\sf RF}(K)].\nn
\end{equation}
Each column vector  targets a corresponding user and is designed to maximize the signal power by coherently combining the corresponding set of data paths under the constraint of nulling inter-cell interference. The design of a single column vector of ${\bf F}_{\sf RF}$  follows the proposed framework of designing ${\bf f}_{\sf RF}$ for the single-user case as summarized in Algorithm \ref{algorithm:2}. Next, substituting  the resultant  ${\bf F}_{\sf RF}$ into (\ref{mmse:1}) gives the corresponding  digital beamformer ${\bf F}_{\sf BB}$. Finally, combining ${\bf F}_{\sf RF}$ and ${\bf F}_{\sf BB}$  yields the Kronecker hybrid beamformer  for the multiuser case.

\subsubsection{CSI Requirement} As reflected in the proposed framework, the  Kronecker analog beamforming requires different levels of CSI for interference nulling (Step~$2$) and signal enhancement (Step~$3$). Specifically, only the AoA of the interference paths are required in Step~$2$ but both AoA and gains of the data paths are needed in Step~$3$. Channel estimation for acquiring the CSI is addressed in Section \ref{sec:VI}.

\subsubsection{Computation Complexity}\label{remark:4}
It is worth pointing out that the proposed Kronecker analog beamforming framework is featured by its low computation complexity compared with the conventional fully digital MMSE beamforming design as elaborated in the following.

\begin{itemize}
\item{Single-user case:} Only the  analog beamformer needs to be computed in this case. Only vector-wise computation is performed in the three steps including  the Fourier based construction in \eqref{in:2},  the signal enhancement  in (\ref{ue:3}), and the final combination of Kronecker factors in \eqref{kd:2}. The resultant  computation complexity is  $O(N)$. In contrast, the conventional fully digital MMSE beamformer requires the  inversion of an  $N\times N$ matrix, which has the  complexity of $O(N^3)$. Such complexity  is too high for massive MIMO systems as $N$ can be as large as several hundred or even thousands. 

\item{Multiuser case:} {For this case, the proposed hybrid beamforming consists of analog and digital beamforming. Following the discussion for the single-user case, the complexity of computing the multiuser analog beamformer is  $O(N \times K)$. Due to dimension reduction by analog beamforming, the subsequent digital beamformer requires the computation complexity of  $O(K^3)$ only. As a result, the total complexity of computing the Kronecker hybrid beamformer is  $O(N \times K) + O(K^3)$, which is much lower than that of the fully digital MMSE beamformer, namely  $O(N^3)$,  when $N$ is large.}
\end{itemize}

\subsubsection{Adaptive Kronecker Factor Allocation} { The proposed Kronecker analog beamforming solution can be further enhanced by adaptively allocating the Kronecker factor according to the transmission environment. Particularly, for the interference-limited scenario where there exist many interference paths, one can choose to cancel only those significant interference such that more Kronecker factors can be saved for signal power enhancement. For the noise-limited case, all the Kronecker factors can be allocated for signal power improvement instead of being wasted on nulling those insignificant interference paths.}

\subsubsection{Antenna Selection} { For the case that the number of BS antenna is a prime number, say 131 antennas, a simple solution is antenna selection where the best 128 antennas are selected according to some selection criterion (e.g., channel quality), and then the Kronecker analog beamforming is performed using the selected 128 antennas. A diversity gain can be attained by antenna selection. Furthermore, it is worth mentioning that the number of antennas of a base station is usually set to be a power of two in the standards such as IEEE 802.11n and  IEEE 802.11ac, fitting the proposed design.}

\section{Channel Estimation for Hybrid Beamforming}\label{sec:VI}
In this section, the channel estimation problem formulated in Section~\ref{sec:III} is solved by designing a two-stage estimation scheme targeting the hybrid beamforming architecture in Fig.~\ref{Fig:2}. First, the AoA of the data/interference paths are estimated by analog beam scanning using the phase array. Next, the data-path gains are estimated using either the analog coherent-combining or analog ZF technique. 

\subsection{Path AoA Estimation}
The key idea of the proposed scheme for path AoA estimation is to construct an \emph{AoA spectrum} by analog beam scanning, which displays power distribution over the AoA range. For ease of exposition, the scheme is first discussed for the simple single-user case without inter-cell interference  and then extended to the general  multiuser case with inter-cell interference. 

\subsubsection{Single-User Case without Inter-cell Interference}
For the current case, leveraging the high spatial resolution of the large-scale array, the observation in a single pilot symbol duration is sufficient for accurate path AoA estimation. Consider the observation from the BS array in an arbitrary pilot symbol duration, which follows from \eqref{aoa:2.1} as 
\begin{align}
{\bf y} = \sum_{\ell=1}^La_\ell{\bf v}{(\Phi_\ell)}x + {\bf n},
\end{align}
where $x$ now denotes the pilot symbol with unit power. The objective of AoA estimation is then to infer the AoA, $\{\Phi_\ell\}$, from $\by$. To this end, define the AoA spectrum as the function  ${\cal F}$: $\mathbb{R}^N \rightarrow \mathbb{R}_+$ given by 
%\begin{align}\label{aoa:1}
%{\cal F}(\Omega) = \frac{|{\bf v}(\Omega)^H{\bf y}|}{\|{\bf v}(\Omega)\|^2}. 
%\end{align}
\begin{align}\label{aoa:1}
{\cal F}(\Omega) = \frac{|{\bf v}(\Omega)^H{\bf y}|}{N}. 
\end{align}
It can be observed that ${\cal F}(\Omega)$ measures the magnitude of the coefficient obtained by projecting $\by$ onto the (beam) steering   vector ${\bf v}(\Omega)$ given in \eqref{sys:2} corresponding to the direction specified by the  angle  $\Omega$.  To derive a closed-form expression for the AoA spectrum, a useful property for the  steering vector ${\bf v}$ is provided  as follows. Given two angles $\Omega, \Phi \in [0,2\pi]$, the \emph{normalized inner product}   between the  steering vectors ${\bf v}(\Omega)$ and ${\bf v}(\Phi)$ is denoted as ${\cal J}$ and  defined as 
\begin{align}
{\cal J}(\Phi - \Omega) = \frac{|{\bf v}(\Omega)^H{\bf v}(\Phi) |}{N}. 
\end{align}
%\begin{align}
%{\cal J}(\Phi - \Omega) = \frac{|{\bf v}(\Omega)^H{\bf v}(\Phi) |}{{\|{\bf v}(\Omega)\|^2}}. 
%\end{align}
\begin{lemma}\label{lemma:1}\emph{
The normalized inner product between the steering vectors ${\bf v}(\Omega)$ and ${\bf v}(\Phi)$ is given as
\begin{align}
{\cal J}(\Phi - \Omega) = \left|{\frac{ {\sf sinc} \({\frac{N}{2}|\Phi - \Omega|}\)}{{\sf sinc} \(\frac{1}{2}|\Phi - \Omega|\)}}\right|.
\end{align}
Given $x \ll 1$, as $N \rightarrow \infty$, 
\begin{equation}
{\cal J}(x)  \leq  \frac{2}{N|x|} + O\left(\frac{1}{N}\right). 
\end{equation}
}
\end{lemma}

\proof
See Appendix \ref{appendix:lemma:1}. 
\endproof

\begin{figure}[tt]
\centering
\includegraphics[width=8cm]{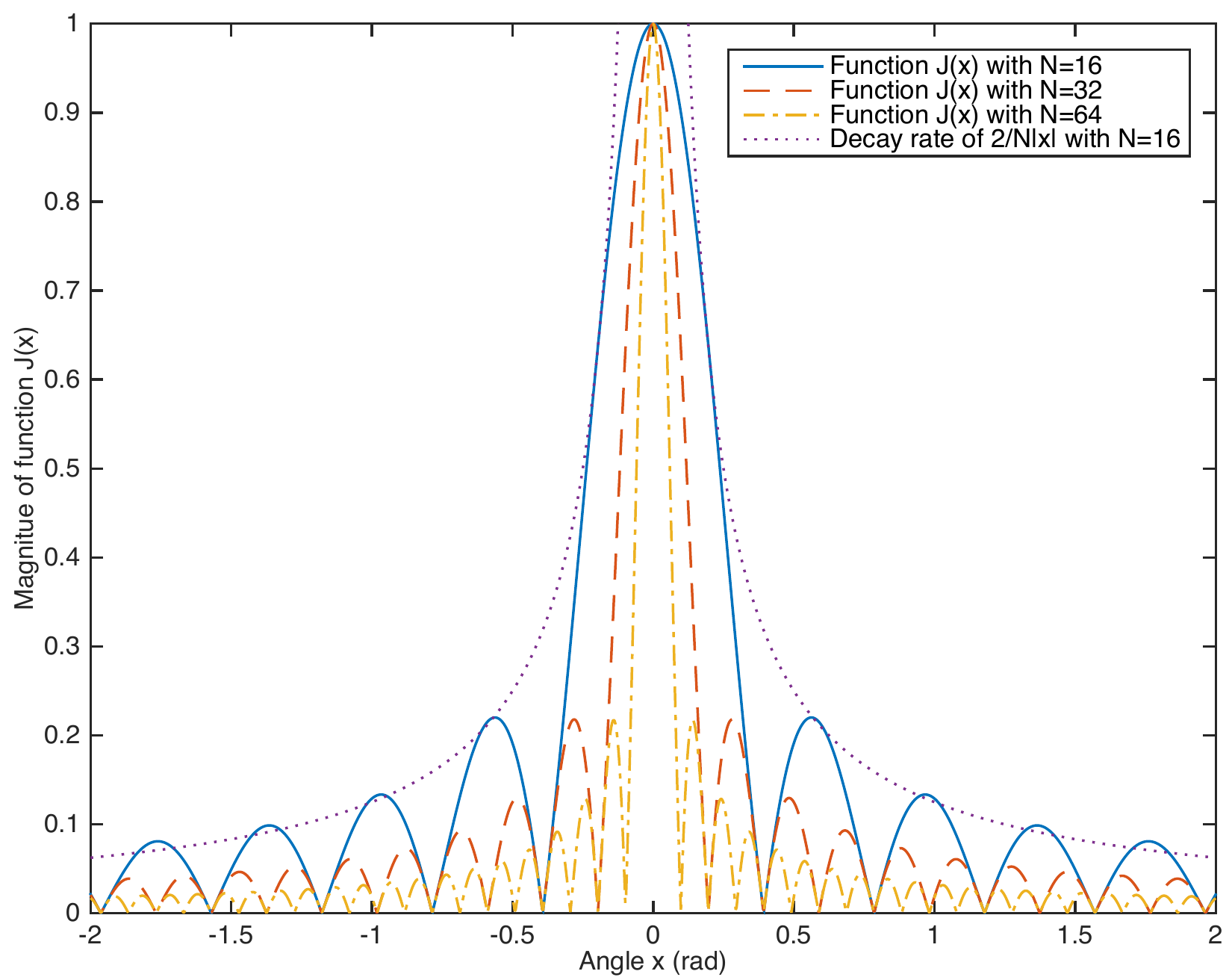}
\caption{The normalized inner product function ${\cal J}(x)$ given in Proposition~\ref{prop:5}.}
\label{Fig:3}
\end{figure}

The function ${\cal J}(x)$ as  plotted  in Fig. \ref{Fig:3} is observed to have a peak at $x = 0$ with magnitude $1$. Moreover, the main lobe has the width of  $\frac{2\pi}{N}$ that becomes narrower as the array size $N$ increases, and the side-lobe diminishes approximately as $O(1/N)$.  Substituting the result in Lemma~\ref{lemma:1} to (\ref{aoa:1}) gives the following proposition.
\begin{proposition}[AoA Spectrum without Interference]\label{prop:5} \emph{For the single-user case without inter-cell interference, 
the AoA spectrum is given as 
\begin{align}
{\cal F}(\Omega) = \left|{\sum_{\ell=1}^L  \tilde a_\ell  {\cal J}(\Phi_\ell - \Omega) + \tilde n}\right|
\end{align}
where $\tilde a_\ell =  {\sf sgn}\left(\frac{\sin\left({\frac{N}{2}(\Phi_{\ell} - \Omega)}\right)}{\sin\left({\frac{1}{2}(\Phi_{\ell} - \Omega)}\right) } \right) a_\ell e^{j\frac{(N-1)(\Phi_\ell - \Omega)}{2}}$ denotes the $\ell$-th effective path gain received by the steering beam, ${\sf sgn}(x)$ is the sign function indicating the sign of $x$, ${\cal J}(\Phi_\ell - \Omega)$ is given in Lemma~\ref{lemma:1} and  $\tilde n = \frac{{\bf v}(\Omega)^H{\bf n}}{N}$ denotes the effective noise observation with zero mean and variance of ${\sf E}\{{\tilde n \tilde n^*}\} = \frac{N_0}{N}$.
}
\end{proposition}
Based on the mentioned property of the function $\mathcal{J}(x)$, the AoA spectrum given in Proposition~\ref{prop:5} comprises $L$ main lobes located at the path AoA $\{\Phi_\ell\}$ with magnitudes $\{|a_\ell|\}$, leading to the proposed procedure for path AoA estimation as follows.

\begin{figure}[tt]
\centering
\includegraphics[width=14cm]{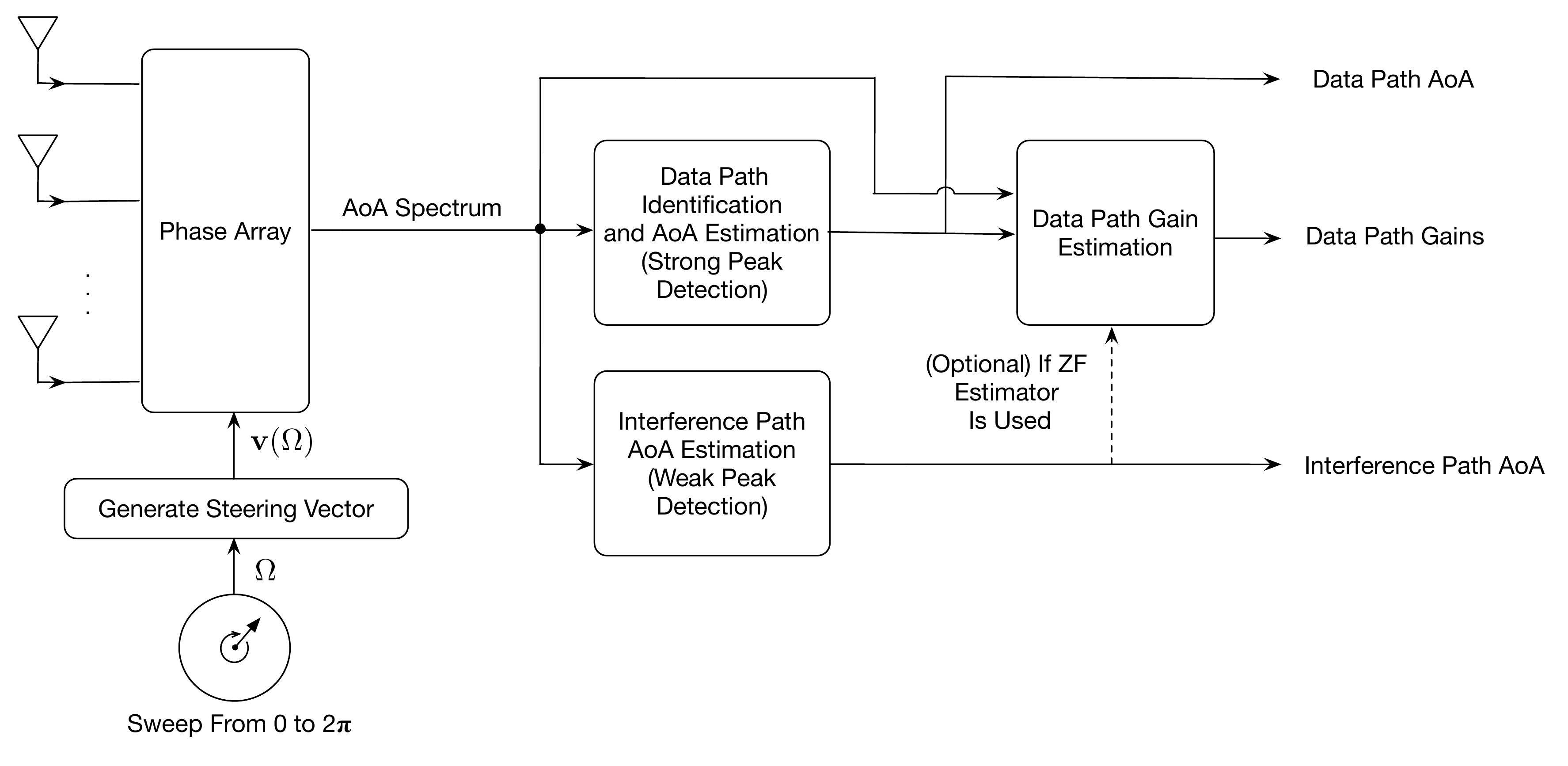}
\caption{The process of channel  estimation for hybrid beamforming.}
\label{Fig:AoA:Estim}
\end{figure}

{\bf Data-Path AoA Estimation:} Essentially, the proposed scheme of path AoA estimation is to detect the peaks in the AoA spectrum. The detailed process  is illustrated in Fig.~\ref{Fig:AoA:Estim}, where the current case without inter-cell interference requires only the upper processing path corresponding to ``strong peak detection". 

\begin{Remark}[Accuracy of the AoA Estimation]\label{remark:8}
\emph{Due to their  side lobes, the set of shifted functions $\{{\cal J}(\Phi_\ell - \Omega)\}$ may interfere with each other and thereby cause inaccuracy in AoA estimation especially in the presence of noise. As the array size $N$ increases, the issue subsides as the side lobes of the functions diminish and the main   lobes become shaper as mentioned. This can be interpreted as that a large-scale array has a high spatial resolution and is capable of resolving  paths even with small AoA separations. Furthermore, it can be observed from Proposition~\ref{prop:5} that the variance of observed  noise  is reduced by the factor of  $\frac{1}{N}$, showing the noise suppression gain of large-scale array.}
\end{Remark}

\begin{Remark}[Parallel Beam Scanning]\label{remark:9}
\emph{As shown in Fig.~\ref{Fig:AoA:Estim}, the implementation of the proposed scheme requires beam scanning of the AoA spectrum over  the  angular range $[0, 2\pi]$.  Hence, the implementation complexity is determined by the \emph{scanning resolution}, defined by $R_{\sf scan} = \frac{2\pi}{N_{\sf sam}}$, where $N_{\sf sam}$ is the required number of samples for constructing  the AoA spectrum. The  scanning process can be accelerated by  \emph{parallel beam scanning} that generates  parallel streams of spectrum samples. Specifically, given $N_{\sf RF}$ RF chains, the phase array, represented by the matrix ${\bf F}_{\sf RF}$, can be divided into $N_{\sf RF}$ subsets of phase shifters, each of which corresponds  to a single column of ${\bf F}_{\sf RF}$ and can be used to implement a single steering vector with a given AoA (see Fig.~\ref{Fig:2}). As a result, the RF chains generate $N_{\sf RF}$ samples of the AoA spectrum each time, reducing the scanning duration by the factor of  $\frac{1}{N_{\sf RF}}$. Sampling the AoA spectrum by beam scanning has the complexity of $O(N)$, which is much lower than the conventional fully digital MMSE channel estimation approach with the complexity of  $O(N^3)$.  { In addition, thanks to the shortened scanning duration and also the low complexity sampling process, a complete AoA range scanning can be done within one pilot sequence duration, hence the resultant training duration equals to the length of the pilot sequence which is relatively short.}
}
\end{Remark}

\begin{Remark}[Comparison with the MUSIC Algorithm]\label{remark:10}
\emph{Both the proposed scheme and the classic \emph{multiple signal classification} (MUSIC) algorithm \cite{schmidt1986multiple} are based on the same principle of scanning over the AoA range. {However, it is difficult to directly  implement the MUSIC algorithm, which targets fully digital processing,  using the hybrid beamforming architecture due to its hardware limitations. Specifically, the algorithm requires the knowledge of the autocorrelation matrix of the full-dimension channel observation, i.e.,  $R_{yy} = {\sf E}\{{\bf yy}^H\}$, which is unavailable for hybrid beamforming due to  the mentioned channel subspace sampling limitation.} Even if such knowledge is available, the key step of the MUSIC algorithm, namely  \emph{singular value decomposition} (SVD) of the matrix $R_{yy}$ with complexity of $O(N^3)$, is impractical for massive MIMO systems since $N$ is too large. In general, the strong capabilities of the large-scale array on noise-and-interference suppression (see Remark~\ref{remark:8}) are sufficient for accurate AoA estimation and  make the use of MUSIC algorithm  less appealing for massive MIMO systems. 
}
\end{Remark}

\subsubsection{Multiuser Case with Inter-cell Interference}\label{Section:MU}

In the presence of multiple users and inter-cell interference, pilot sequences are considered for suppressing both intra-cell and inter-cell interference in the estimation process. The sequences also serve the purpose of identifying the data paths for different users. 

\vspace{5pt}
\noindent\underline{a) Data-Path AoA Estimation} \newline
Consider the AoA estimation for the $k$-th user, the observation matrix in \eqref{aoa:2.1} is multiplied by the corresponding pilot sequence $\bx_k$ with normalization, yielding

\begin{equation}\label{Eq:Training:MU}
\tilde{\by}_k = \frac{{\bf Y}\bx_k}{\|\bx_k\|^2} = \sqrt{Z}{\bf g}_k + \sum_{\ell=1}^M\gamma_\ell{\bf h}_\ell  +   \tilde{\bf n}, 
\end{equation}
where $\{\gamma_\ell\}$ are i.i.d. random variables with zero mean and unit variance and the noise vector comprises of i.i.d. $\mathcal{CN}(0, N_0)$ elements. The intra-cell interference is eliminated in \eref{Eq:Training:MU} due to the orthogonality of pilot sequences used in the same cell. {Note that the pilot sequence coherent combining is indeed performed in the digital baseband after the analog beam scanning process. We switch the presentation order here as it can simplify the expression without any violation of the mathematical consistency and also facilitate to establish a clear link to the single-user case.}

Next, following \eqref{aoa:1}, define the AoA spectrum for the $k$-th user in the current case  as 
\begin{align}\label{Eq:Spectrum:Mu}
{\cal F}_k(\Omega) = \frac{|{\bf v}(\Omega)^H{\tilde{\by}_k}|}{N}. 
\end{align}
%\begin{align}\label{Eq:Spectrum:Mu}
%{\cal F}_k(\Omega) = \frac{|{\bf v}(\Omega)^H{\tilde{\by}_k}|}{\sqrt{Z}\|{\bf v}(\Omega)\|^2}. 
%\end{align}
The spectrum can be obtained using the same method of beam scanning as in the preceding single-user case, yielding the following result.
\begin{proposition}[AoA Spectrum with Inter-cell Interference]\label{prop:6}\emph{
In the presence of inter-cell interference, the AoA spectrum  for the $k$-th user is given by 
\begin{align}\label{aoa_spectrum2}
{\cal F}_k(\Omega) = \left|{\sum_{\ell=1}^L \tilde a_{k\ell}  {\cal J}(\Phi_{k\ell} - \Omega) + \frac{1}{\sqrt{Z}}\sum_{n=1}^M  \gamma_{n} \tilde \beta_n  {\cal J}(\Theta_n - \Omega) + \tilde n_k}\right|,
\end{align}
where $\tilde a_{k\ell} = {\sf sgn}\left(\frac{\sin\left({\frac{N}{2}(\Phi_{k\ell} - \Omega)}\right)}{\sin\left({\frac{1}{2}(\Phi_{k\ell} - \Omega)}\right) } \right) a_{k\ell}e^{j\frac{(N-1)(\Phi_{k\ell} - \Omega)}{2}}$ and 
$\tilde \beta_n = {\sf sgn}\left(\frac{\sin\left({\frac{N}{2}(\Theta_n - \Omega)}\right)}{\sin\left({\frac{1}{2}(\Theta_n - \Omega)}\right) } \right) \beta_n e^{j\frac{(N-1)(\Theta_n - \Omega)}{2}}$ 
are the effective data and interference path gains respectively, $\tilde n_k = \frac{{\bf v}(\Omega)^H}{N}{\bf N}\frac{{\bf x}_k}{\|x_k\|^2}$ denotes the effective noise with zero mean and variance of $\frac{N_0}{Z N}$.
%$\gamma_k = {\bf x}_k^T\frac{{\bf x}_k}{\|{\bf x}_k\|}$ is the coherent combining gain for the desired user $k$, $\gamma'_{\ell} = {\bf s}_\ell^T\frac{{\bf x}_k}{\|{\bf x}_k\|}$ denotes the combining gain for the $\ell$-th inter-cell interfering ray, which is a random variable with zero mean and unit variance as indicated in Lemma \ref{lemma:2}, and $\tilde n_k = \frac{1}{N}{\bf v}(\Omega)^H{\bf N}\frac{{\bf x}_k}{\|x_k\|}$ denotes the effective noise observation with zero mean and variance of $\frac{N_0}{N}$. The intra-cell interference terms are canceled due to ${\bf x}_k^T{\bf x}_j = 0$, $\forall k\neq j$.
}
\end{proposition}

\begin{figure*}[tt]
  \centering
  \subfigure[AoA Spectrum  with $Z = 1$]{\label{fig:4a}\includegraphics[width=0.42\textwidth]{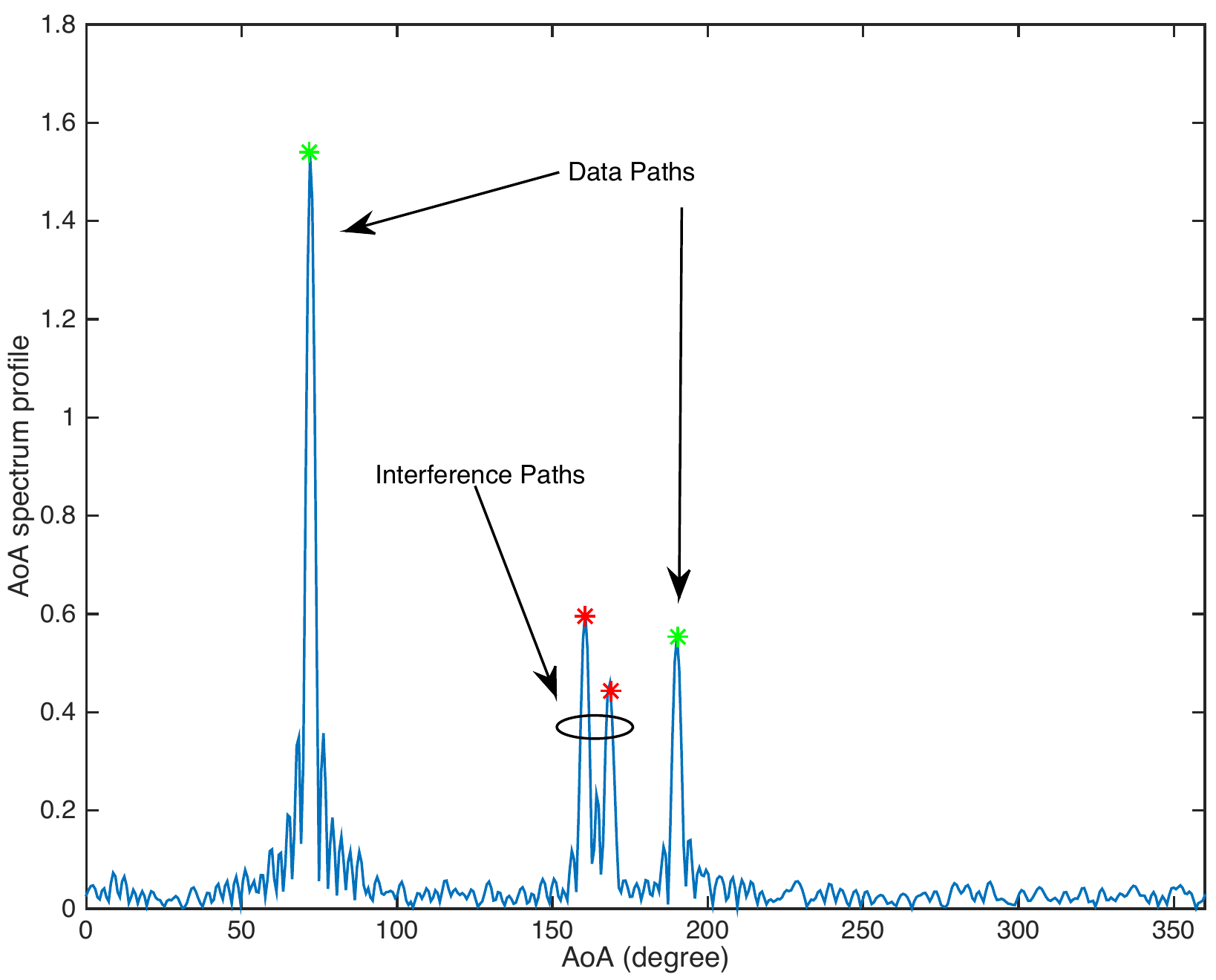}}
  \hspace{0.2in}
  \subfigure[AoA Spectrum  with $Z =10$]{\label{fig:4b}\includegraphics[width=0.42\textwidth]{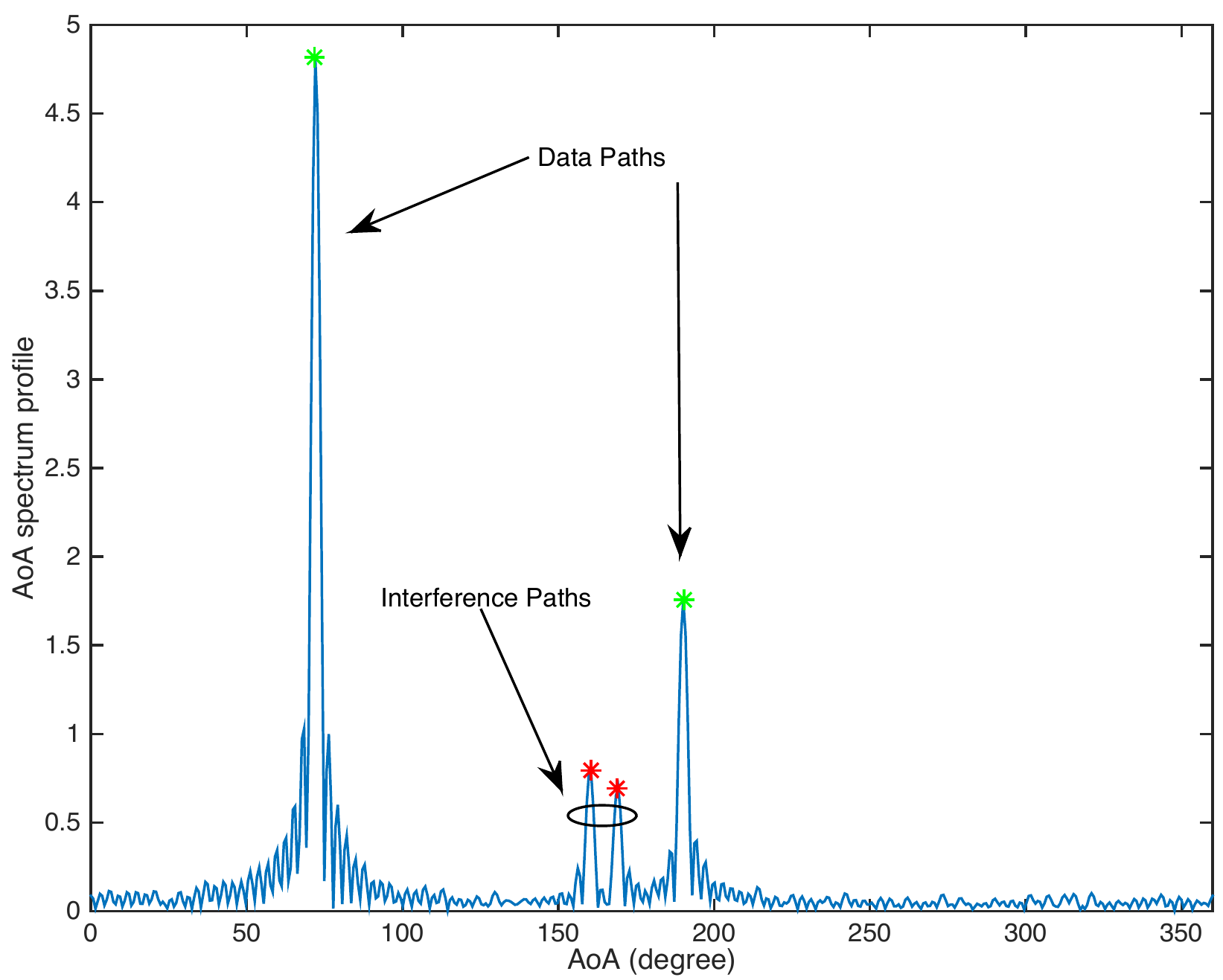}}
  \caption{The effect of the pilot sequence length $Z$ on suppressing inter-cell interference in the AoA spectrum with $N = 128$, $K=1$, $M=2$ and $L=2$.}
  \label{Fig:4}
\end{figure*}

In the AoA spectrum in \eqref{aoa_spectrum2}, the second summation represents inter-cell interference for  data-path AoA estimation, which   arises from pilot contamination. Note that the interference magnitude is suppressed by the factor $1/\sqrt{Z}$ due to the use of pilot sequences as also illustrated  in Fig.~\ref{Fig:4}. 

{\bf Data-Path AoA Estimation:} The process is identical to the single-user counterpart, involving peak detection of the AoA spectrum,  where the spectrum for the current case refers to ${\cal F}_k(\Omega)$ in Proposition~\ref{prop:6}. Note that given sufficient long pilot sequences, ${\cal F}_k(\Omega)$ comprises of strong and weak  peaks (reduced by the factor of $1/\sqrt{Z})$ corresponding to data and interference paths, respectively. For detecting the data paths, it is necessary to separate the strong peaks from the weak ones by e.g., applying a threshold.

Next, the dependence of pilot contamination on various system parameters are quantified as follows. Define  \emph{pilot-contamination level}, denoted as $\eta$, as the maximum magnitude of interference in the AoA spectrum  ${\cal F}_k(\Omega)$ (see Proposition~\ref{prop:6}) at the AoA of a data path: 
\begin{align}
\eta &=\max_{\ell} \frac{1}{\sqrt{Z}} \left| \sum_{n = 1}^M \gamma_n \tilde \beta_n {\cal J}(\Theta_n -  \Phi_\ell) \right|.  \nn\\
&\leq  \frac{\sum_{n = 1}^M |\gamma_n \tilde \beta_n|}{\sqrt{Z}}  \max_{n, \ell} {\cal J}(\Theta_n -  \Phi_\ell).  \nn
\end{align}
Using the result in Lemma~\ref{lemma:1}, $\eta$ can be bounded as follows. 

\begin{proposition}[Effect of Pilot Contamination]\label{prop:PC}\emph{For data-path AoA estimation, the pilot-contamination level can be bounded as 
\begin{equation} 
\eta  \leq  \frac{a}{\sqrt{Z} N  \min\limits_{n, \ell} |\Theta_n -  \Phi_\ell|},  \qquad N \rightarrow \infty,  \label{Eq:PilotC}
\end{equation}
where $a$ is a constant. 
}
\end{proposition}
The result shows the decay rate of $\eta$ with increasing array size $N$, pilot-sequence length $Z$ or the minimum  spatial separation between data and interference paths,  $\min\limits_{n, \ell} |\Theta_n -  \Phi_\ell|$.   It was first observed in \cite{YinGesbert:CoordinatedApproachLargeScaleMIMO:2013} that the interference to channel estimation due to pilot contamination vanishes as $N$ increases provided that  the interference and data paths are spatially separated, namely that $\min\limits_{n, \ell} |\Theta_n -  \Phi_\ell| >0$. Align with this observation, the result in \eqref{Eq:PilotC} is more elaborate and quantifies the scaling law of the interference magnitude with the system parameters. 

\vspace{5pt}
\noindent\underline{b) Interference-Path AoA Estimation} \newline
Estimating the AoA of interference paths is to simply detect the weak peaks in one of the $K$   AoA spectrums  in \eqref{aoa_spectrum2} where the weak peaks are farthest  from the strong  ones to reduce their interference. The process corresponds to the lower processing path in Fig. \ref{Fig:AoA:Estim}. For the case with insufficient AoA separations between data and interference paths, the side-lobes of strong data-path peaks in the AoA spectrum can cause severe interference to detecting the weak interference-path peaks. To address this issue, the performance of the estimation of  the interference-path AoA can be improved by \emph{decision feedback} following the estimation of the AoA and gains of the data paths\footnote{The gain estimation of the data paths can be conducted using the analog coherent-combining estimator as shown in the sequel.}. 
Specifically, those data path components can be removed  from their corresponding  AoA spectrums in  \eqref{aoa_spectrum2} by subtraction prior to estimating the interference-path AoA.  The potential issue of this method is error propagation due to inaccurate data-path estimation. The issue can be mitigated by averaging the $K$ AoA spectrums after the said subtraction to suppress the residual data-path components as well as noise. 

\subsection{Estimation of Data-Path Gains}\label{path_gain_esti}
In this section, given the estimated data-path AoA $\{\Phi_\ell\}$, the schemes for estimating the corresponding path gains $\{a_\ell\}$ are discussed. To simplify the exposition, consider the single-user case with inter-cell interference and the observation in an arbitrary pilot symbol duration can be reduced from \eqref{aoa:2.1}  as 
\begin{align}\label{PGE:1}
{\bf y} = \sum_{\ell=1}^La_\ell{\bf v}{(\Phi_\ell)}x + \sum_{n=1}^M\beta_n{\bf v}{(\Theta_n)}s_n + {\bf n},
\end{align}
where $x$ and $s_n$ are the pilot symbols transmitted by the intended user and the $n$-th interfering user, respectively. Without loss of generality, set $x=1$. The extension to the multiuser case is similar to that for path AoA estimation in Section~\ref{Section:MU}. 

The proposed scheme of estimating the gain of a particular data path is to steer the analog beamformer such that its output is a scaled version of the gain. Let the analog beamformer for estimating the $\ell$-th path gain be denoted as $\bff_{\sf RF}(\ell)$. We propose two designs of the beamformer, called the analog \emph{coherent-combining} and analog \emph{zero-forcing} (ZF) methods, such that  the beamformer output is 
\begin{equation}
\hat{y}_\ell = [\bf f_{\sf RF}(\ell)]^H {\bf y}\approx c_\ell a_\ell,
\end{equation}
where $c_\ell$ is a constant. By scaling in the digital domain, the estimated gain is obtained as $\hat{a}_\ell = \hat{y}_\ell/c_\ell$. 

\subsubsection{Gain Estimation by Analog Coherent Combining}
Consider gain  estimation using the analog coherent-combining  beamformer. The design, $\bff_{\sf RF}(\ell) = {\bf v}(\Phi_\ell)$, coherently combines the outputs of all antennas due to the signal from the $\ell$-th data path. With $c_\ell = N$, the estimated gain is 
\begin{equation}\label{GE_CC}
\hat{a}_\ell^{\sf cc}  = a_\ell + \frac{1}{N} \sum_{m\neq \ell}^La_m{\bf v}{(\Phi_\ell)}^H{\bf v}{(\Phi_m)} + \frac{1}{N}\sum_{n=1}^M\beta_n{\bf v}{(\Phi_\ell)}^H{\bf v}{(\Theta_n)}s_n + \frac{1}{\sqrt{N}} \hat{n},
\end{equation}
where the effective noise $\hat{n}$ is a $\mathcal{CN}(0, N_0)$ random variable. 

\begin{proposition}\label{Prop:PathEst:CC}\emph{
For data-path gain estimation using the coherent-combining method, the estimation error $|\hat{a}_\ell^{\sf cc} - a_\ell|$ can be bounded as 
\begin{align}\label{exp:PathEst:CC}
|\hat{a}_\ell^{\sf cc} - a_\ell| \leq \frac{2\alpha_{\max} (L+M-1)}{\Psi_{\sf min} N}  +O\l(\frac{1}{N}\r),
\end{align}
{ where $\alpha_{\max}$ is the maximum of the magnitudes of all path gains, namely the maximum element of the set  $\{|a_\ell|, |\beta_n|\}$ for all $\ell$ and $n$, and $\Psi_{\sf min}$ is the minimum AoA separation between any pair of paths, namely the minimum element of the set  $\{|\Phi_\ell - \Phi_m|, |\Phi_\ell - \Theta_n|\}$ for all $\ell$, $m$ and $n$.}
}
\end{proposition}

\proof
See Appendix \ref{appendix:Prop:PathEst:CC}.
\endproof
The result shows that the path gain estimation is highly accurate for a large-scale array ($N \rightarrow\infty$). Moreover, the accuracy increases as the path spatial separations grow.  

\subsubsection{Gain Estimation by Analog ZF}
Consider gain  estimation using the analog ZF beamformer. The design requires prior estimation of the AoA of all  paths, namely $\{ \Phi_{\ell}\}\cup \{\Theta_n\}$. Given the AoA, the ZF beamformer for estimating the path gain $a_\ell$ is designed to null all other path components in 
the observation in \eqref{PGE:1} except for that corresponding to the $\ell$-th data path. Assuming perfect nulling, the only resource for estimation error is noise. To maximize the SNR, the problem of analog ZF beamformer  design can be formulated as follows: 
\begin{equation}({\bf P7})\qquad 
\begin{aligned}
\mathop {\max }\limits_{ {\bf f}_{\sf ZF}(\ell) } \; & |{\bf f}_{\sf ZF}^H(\ell){\bf v}(\Phi_\ell)|^2 \\
{\textmd{s.t.}}\;\; & |{\bf f}_{\sf ZF}^H(\ell){\bf v}(\Phi_m)|^2 = 0, \;|{\bf f}_{\sf ZF}^H(\ell){\bf v}(\Theta_n)|^2 = 0, \;\forall m\neq \ell, \;\forall n,\\ 
&|[{\bf f}_{\sf ZF}(\ell)]_{i}| = 1,\;\forall i.
\end{aligned}
%\end{aligned}\nn
\label{P:7}
\end{equation}
The problem can be solved using the proposed Kronecker analog beamforming framework as summarized in  Algorithm \ref{algorithm:2}. Given the ZF beamformer ${\bf f}_{\sf ZF}(\ell)$, the estimated path gain is given as 
\begin{equation}\label{Eq:GainEst:ZF}
\hat{a}_\ell^{\sf zf} = a_\ell +  \frac{{\bf f}_{{\sf ZF}}^H(\ell){\bf n}}{{\bf f}_{{\sf ZF}}^H(\ell){\bf v}(\Phi_\ell)}. 
\end{equation}

\subsubsection{Estimator Comparison}\label{sec:Estimator_Comparison}
The two proposed  path-gain estimation schemes have their own pros and cons. The scheme with coherent combining beamforming maximizes the SNR but fails to suppress interference. In contrast, the other scheme with ZF beamforming nulls interference at the cost of a smaller SNR. The insight into the relative performance of the two schemes can be obtained by considering the simple  case  with single data and single interference  paths ($K=1, L=1$,  $M=1$) and the array size $N$ being  even.  If the data and interference paths are poorly separated in space (i.e., $|\Phi - \Theta|$ is small enough),   it can be shown that ${\bf f}_{{\sf ZF}}^H{\bf v}(\Phi) \approx \frac{-j(\Phi - \Theta)}{2}N$ and the noise in \eqref{Eq:GainEst:ZF} has the variance approximately equal to $\frac{4N_0}{|\Phi - \Theta|^2N}$. Then the ratio between the expected  estimation errors can be written as ({see Appendix \ref{appendix:C}}) 
\begin{align}\label{esti_comp}
\frac{{\sf E}[|\hat a^{\sf cc}_\ell  - a_\ell|]}{{\sf E}[|\hat a^{\sf zf}_\ell  - a_\ell|]} \approx \frac{\rho}{\sqrt{N}} + \frac{|\Phi - \Theta|}{2},
\end{align}
where $\rho = \beta/\sqrt{N_0}$ denotes the ratio between the interference path gain and the noise power. The result suggests that when $\rho$ is large (strong inter-cell interference), the estimator based on analog ZF achieves better performance than that based on analog coherent-combining as the gain of interference nulling outweighs the the resultant loss of SNR.  The result also indicates  that the estimator based on analog ZF  becomes less appealing when the AoA separation  between the interference and data paths is small. For this case, nulling the interfering path may result in significant noise enhancement. Last, as   $N$ increases, the performance gap between the two estimators  diminishes as the estimation errors for both designs converge to zero. 

\begin{Remark}
\emph{The proposed two-stage channel estimation scheme enjoys a significant practical interest due to two main reasons. One is that it involves only simple beam scanning and scalar gain estimation with a low computation complexity of $O(N)$. On the other hand, in mmWave communications, interference paths largely depend on static blockage objects in the environment (e.g., buildings). Thus, the channel variation is slow, making the channel tracking easy.}
\end{Remark}

\section{Simulation Results}
In this section, the performance of the proposed Kronecker analog/hybrid beamformer and the matching channel estimation approach will be evaluated under various system parameter settings.
The simulation parameters are set  as follows unless specified otherwise. The array size  $N=128$, the number of data paths per user $L=2$, the number of interference paths $M=2$, the number of users $K=4$. All path gains, $\{a_{k\ell}, \beta_n\}$,  follow i.i.d. ${\cal CN}(0, 1)$ distributions. The transmit SNRs for intended users, denoted as $\rho_U$, and interfering users, denoted as $\rho_I$,  are set as $0$ dB by default.  The AoA of all paths follow i.i.d. uniform distributions in the range of $[0, 2\pi]$.

\subsection{Kronecker Analog Beamforming for the Single-user Case}
\begin{figure*}[tt]
  \centering
  \subfigure[Effect of the  transmit  SNR of intended users]{\label{fig:6a}\includegraphics[width=0.42\textwidth]{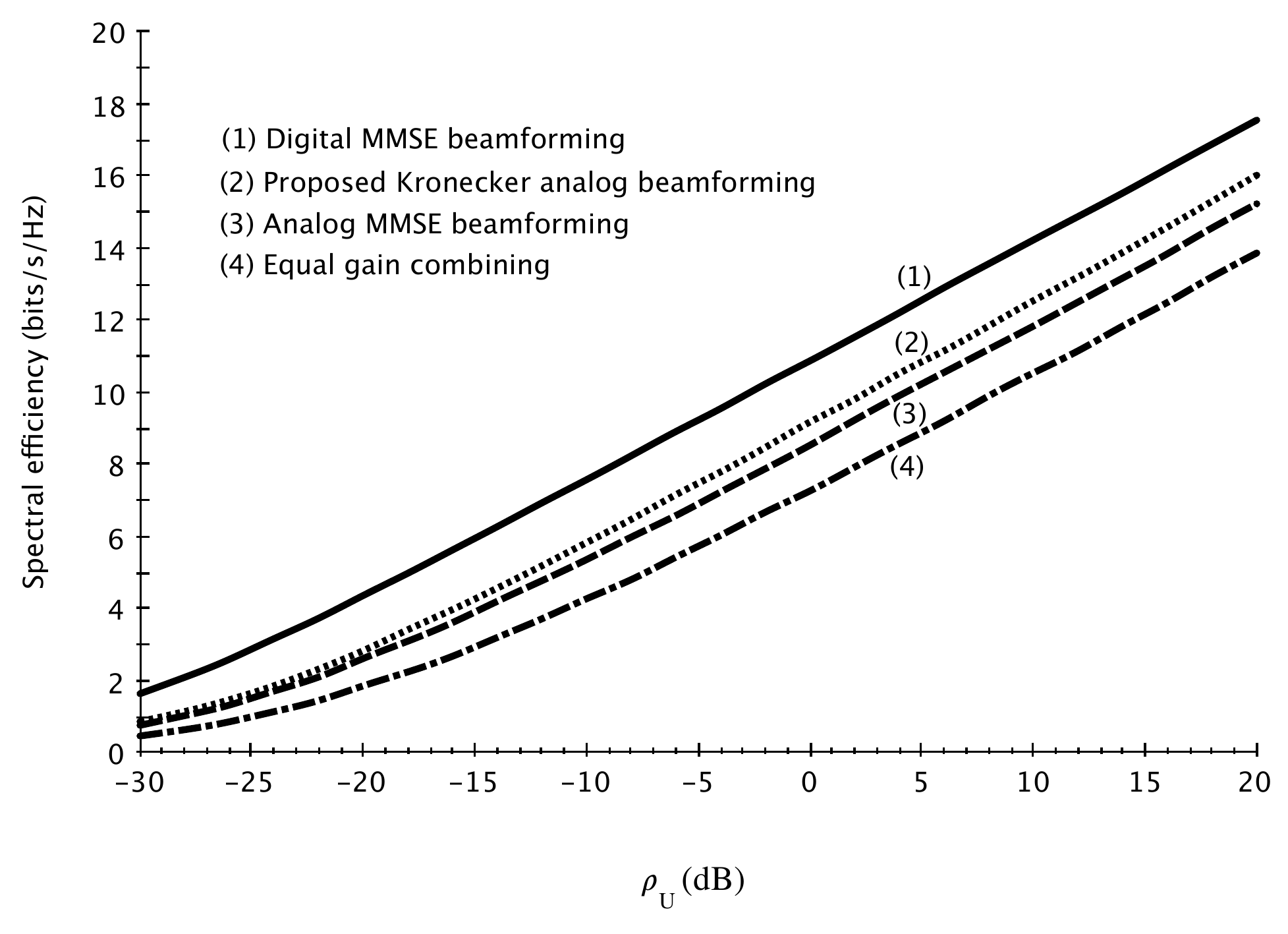}}
  \hspace{0.35in}
  \subfigure[Effect of the transmit  SNR of interfering users]{\label{fig:6b}\includegraphics[width=0.42\textwidth]{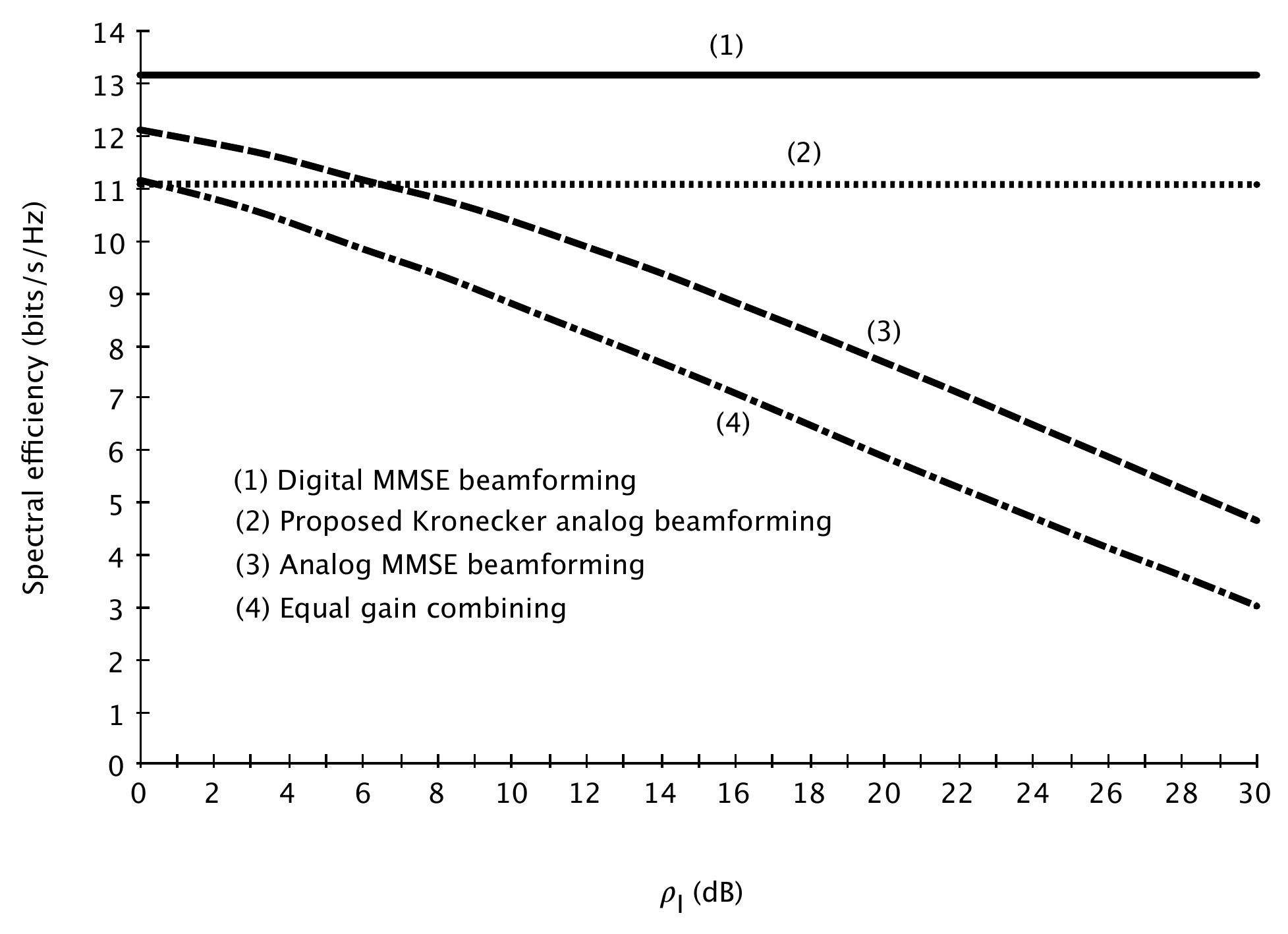}}
  \caption{Comparison of the spectral efficiencies of  the proposed Kronecker analog beamforming and existing digital/analog beamforming designs for the single-user system with inter-cell interference.}
  \label{Fig:6}
\end{figure*}

Consider the single-user case in the presence of inter-cell interference and only a single RF chain is used in the hybrid-beamforming architecture.  In Fig. \ref{Fig:6}, the spectral efficiency of the proposed Kronecker analog beamformer is compared with those of the optimal fully digital MMSE beamformer and two existing designs of analog beamformer. One is the well-known design of \emph{equal-gain combining} where the analog beamforming vector comprises the element phases of the  array observation vector of  the intended user channel (see e.g., \cite{love2003equal}). The  other  is  called \emph{analog  MMSE beamformer}  which approximates the fully digital counterpart by taking only the phases of its coefficients \cite{RobustCL2016}.  
%The   interfering  users' transmit SNR is fixed at  $\rho_I = 10$ dB.  
In  Fig. \ref{fig:6a}, the performances of different beamforming designs are compared with a varying intended users' transmit SNR. It is observed that the proposed design achieves a lower spectral efficiency than the optimal fully digital MMSE design due to the analog  hardware  constraints but outperforms other analog beamforming designs, where the performance gaps are fixed at relatively high $\rho_U$ regime given fixed $\rho_I$. 
In Fig. \ref{fig:6b}, the comparison is over a varying interfering users' transmit SNR. In the weak interference regime, the proposed Kronecker analog beamforming is outperformed by the analog MMSE design due to the noise enhancement effect of the former. For other regimes, the proposed design surpasses the other analog designs. In particular, by analog interference nulling, the spectral efficiency of the Kronecker analog beamforming is independent with the interference level while those of other analog designs decrease as the interference power increases. 

\begin{figure}[tt]
\centering
\includegraphics[width=0.5\textwidth]{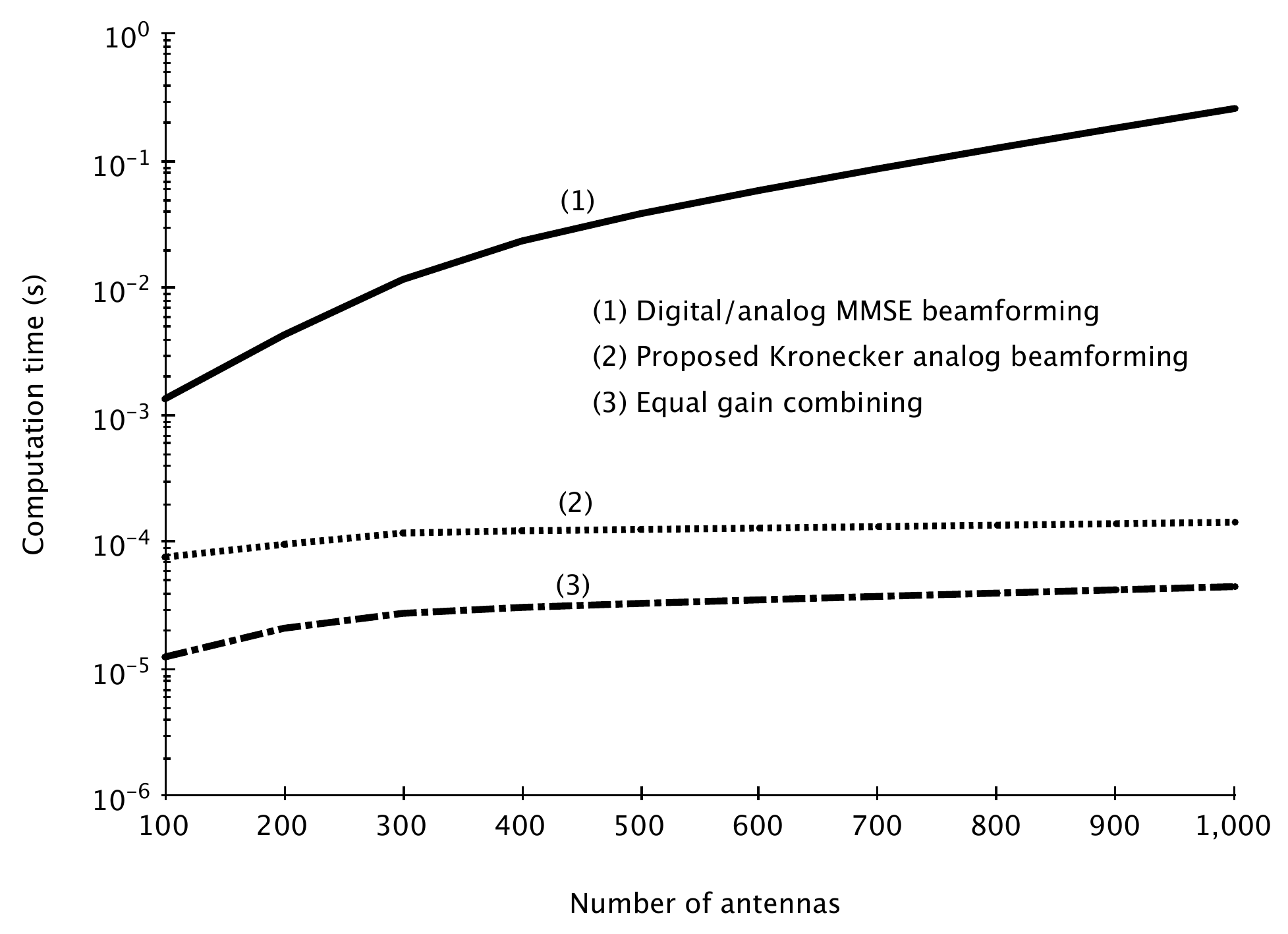}
\caption{Comparison of the computation times of  the proposed Kronecker analog beamforming and the existing digital/analog beamforming designs for the single-user system with inter-cell interference.}
\label{Fig:7}
\end{figure}

In Fig. \ref{Fig:7}, the computation time of the proposed Kronecker analog beamforming is compared with those of other digital/analog designs considered in Fig. \ref{Fig:6}, which are obtained by MATLAB simulations. One can see that the proposed design has a higher complexity than the simple equal-gain combining design but lower complexity than the digital and analog MMSE beamforming. The complexities of the proposed and equal-gain combining designs are both $O(N)$ and thus less sensitive to the variation of the array size $N$ than those of the two MMSE designs which are both $O(N^3)$. The same observations also apply to the multiuser case. 

\subsection{Kronecker Hybrid Beamforming for Multiuser Case}
\begin{figure*}[tt]
  \centering
  \subfigure[Effect of the transmit  SNR of the intended users]{\label{fig:8a}\includegraphics[width=0.42\textwidth]{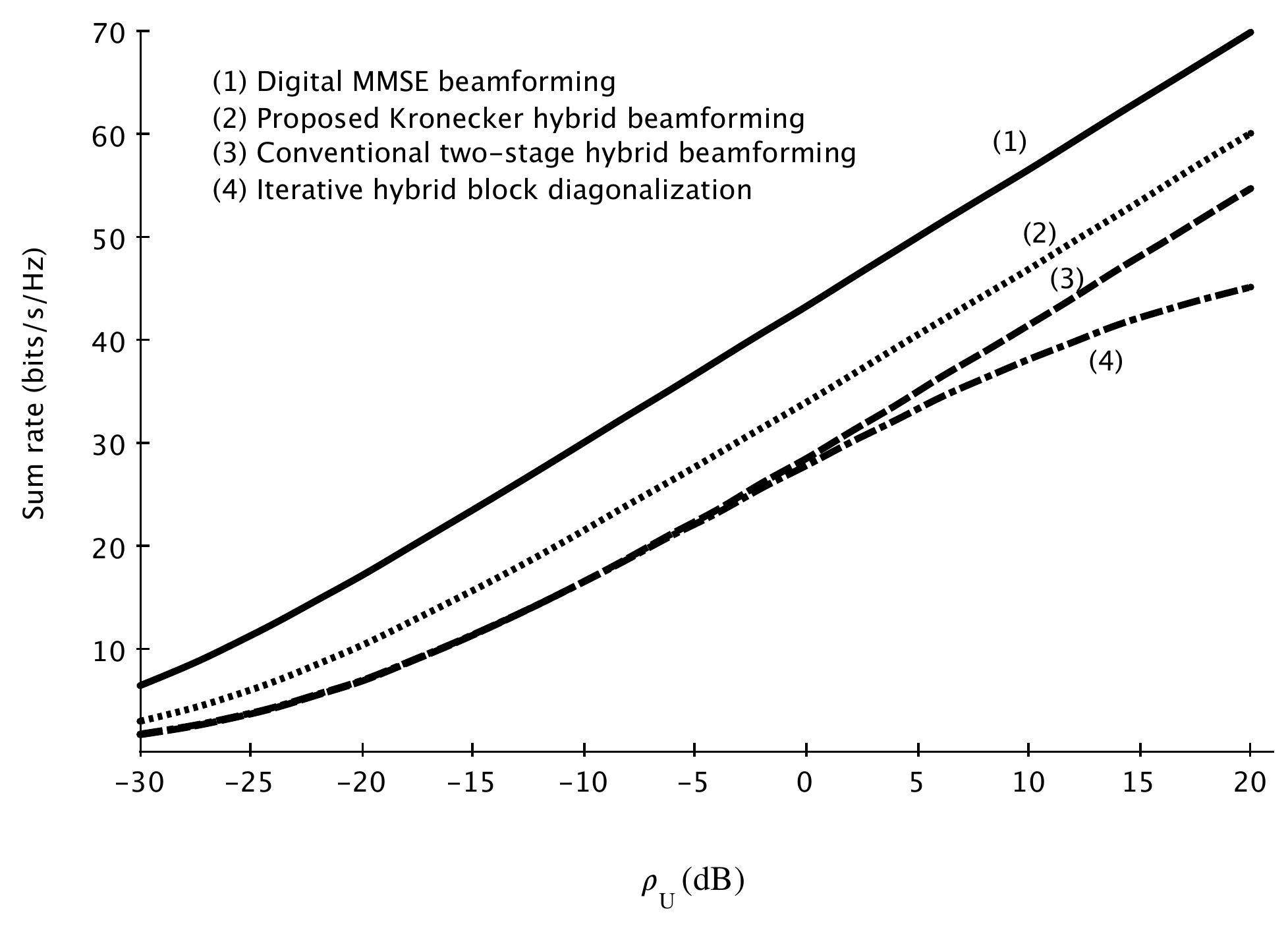}}
  \hspace{0.35in}
  \subfigure[Effect of the  transmit SNR of the interfering users]{\label{fig:8b}\includegraphics[width=0.42\textwidth]{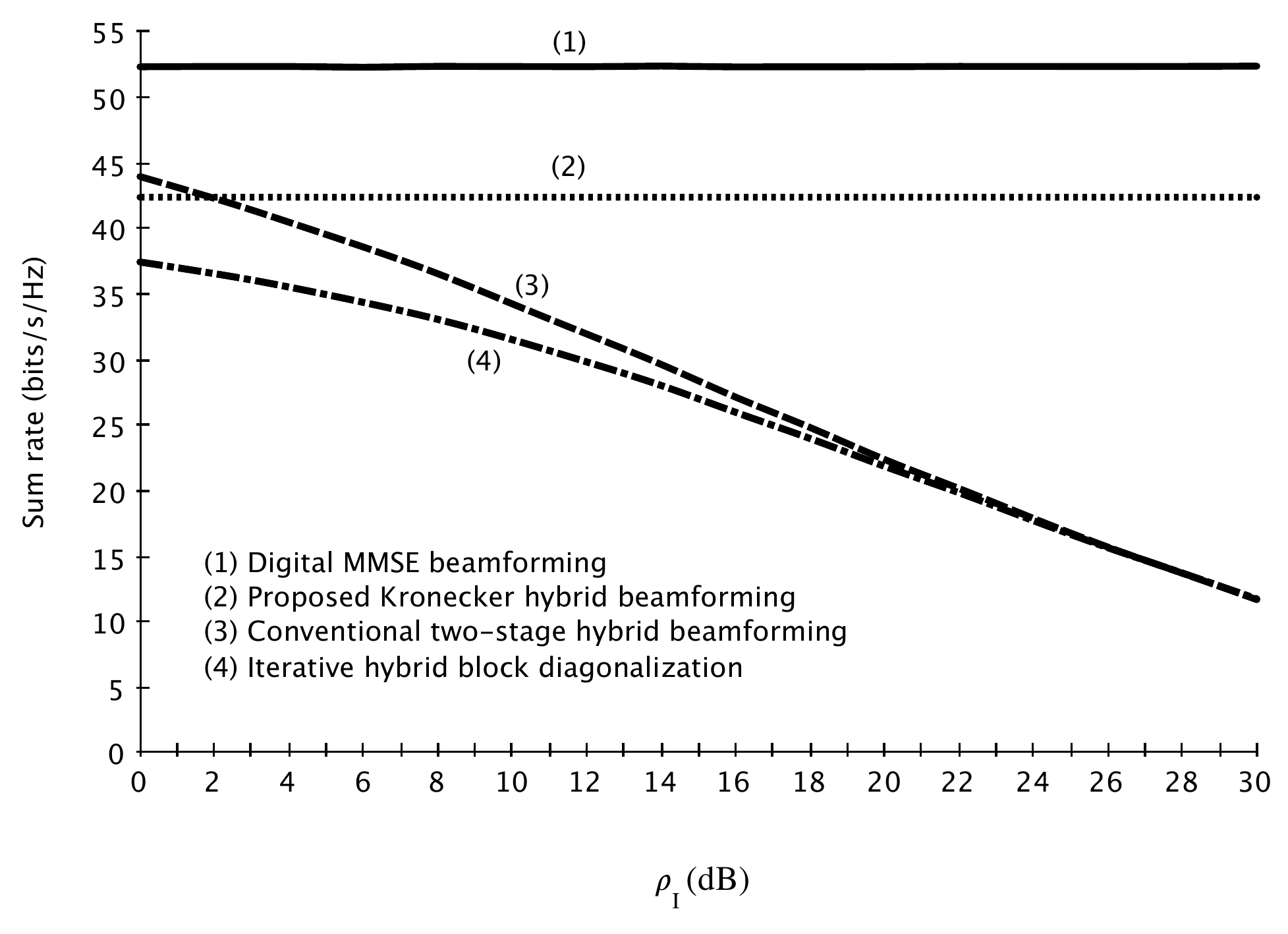}}
  \caption{Comparison of the achievable sum rate  between the proposed Kronecker hybrid beamforming design and the existing designs for the multiuser system with inter-cell interference.}
  \label{Fig:8}
\end{figure*}

{Consider the multiuser case with the number of RF chains $N_{\sf RF} = K = 4$, the sum rate performance of the proposed Kronecker hybrid beamformer is evaluated  in Fig. \ref{Fig:8} via comparison with several existing hybrid beamforming designs. In particular, the \emph{iterative hybrid block diagonalization} design proposed in \cite{RajashekarTWC2016} attempts to minimize the Euclidean distance between the hybrid beamformer and a proposed digital block diagonalization beamformer using a iterative matrix decomposition method. By only approximation, this design is incapable of  nulling  intra-cell and inter-cell interference.  Furthermore, the  \emph{two-stage hybrid beamforming} design proposed in \cite{alkhateeb2015limited} consists of an analog beam steering followed by a  small-scale digital ZF beamformer. Due to the limited number of RF chains and the  resultant issue  of channel subspace sampling limitation, the  low-dimension digital ZF can suppress intra-cell interference but lacks additional DoF for nulling inter-cell interference.  Several observations can be made from  Fig. \ref{Fig:8}. From Fig. \ref{fig:8a}, one can see that the proposed Kronecker hybrid beamforming outperforms other hybrid beamforming designs and achieves a sum rate about $80\%$ of that achieved by the optimal fully digital MMSE beamforming which has much higher complexity in both computation and hardware. Note that the sum rate of the \emph{iterative hybrid block diagonalization} saturates as the intended users' SNR increases due to the existence of intra-cell interference. Next, it can be observed from Fig. \ref{fig:8b} that by analog interference cancellation, the proposed Kronecker hybrid beamforming design is robust against the increase in inter-cell interference power while it incurs increasing rate loss for other hybrid beamforming designs. }

\subsection{Channel Estimation}
\begin{figure*}[tt]
  \centering
  \subfigure[Effect of the number of antennas]{\label{fig:5a}\includegraphics[width=0.42\textwidth]{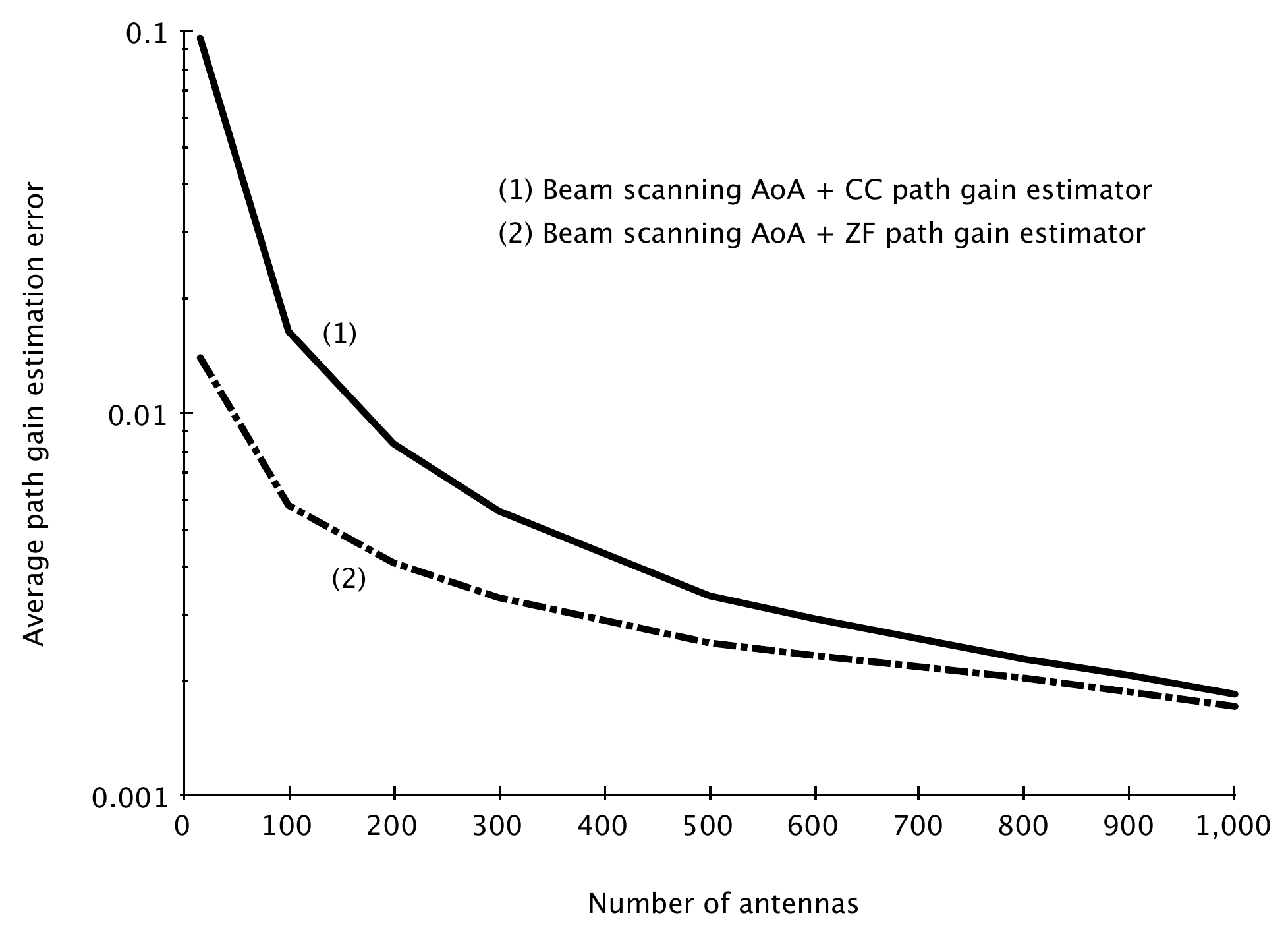}}
  \hspace{0.35in}
  \subfigure[Effect of the transmit SNR of interfering users and the AoA separation between paths]{\label{fig:5b}\includegraphics[width=0.42\textwidth]{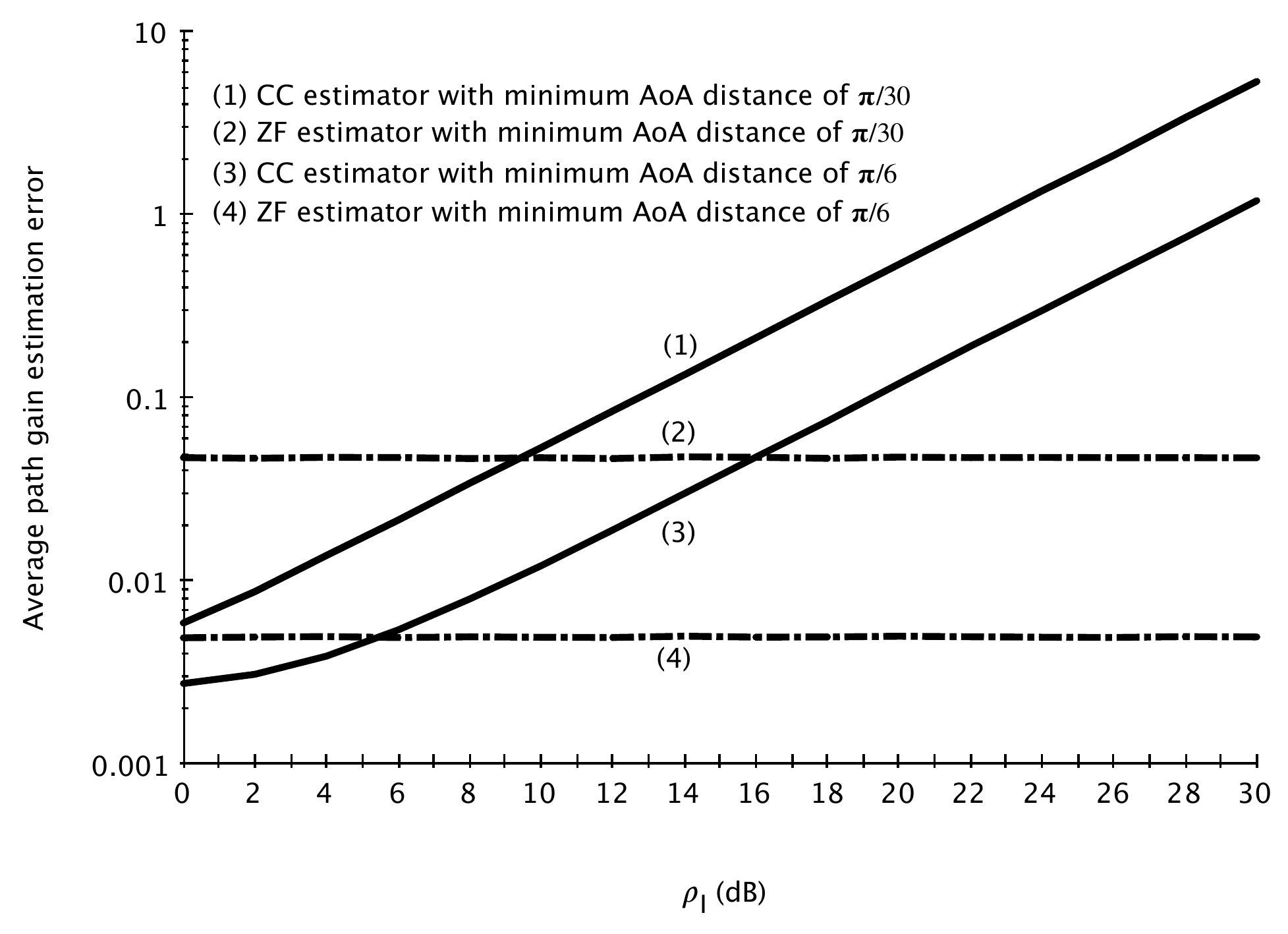}}
  \caption{Performance of the proposed two-stage channel estimation scheme.}
  \label{Fig:5}
\end{figure*}

The performance of the proposed beam-scanning path AoA estimation plus beam-steering data-path gain estimation with analog coherent combining (CC) or analog ZF is evaluated  in Fig. \ref{Fig:5}.  Fig. \ref{fig:5a} shows the average estimation error with a varying  array size $N$. The analog ZF scheme is observed to outperform the analog CC scheme but their performances converge as $N$ increases. The reason is that coherent combining using a large-scale antenna array also suppresses interference thanks to the high spatial resolution. Next, the comparison in  Fig. \ref{fig:5b} is for a varying transmit SNR of interfering users and different minimum spatial separations between propagation paths. One can observe that in the weak interference regime, enhancing the SNR is preferred and thus the CC scheme outperforms the ZF scheme. However, the performance of the CC scheme degrades rapidly as the interference power increases. As a result, the ZF scheme is preferred in the strong interference regime. Furthermore, it is found that increasing the minimum path spatial separation improves both estimators' performance significantly which aligns with our previous analytical results.  

\section{Extensions}
In this section, several possible extensions of the proposed framework will be briefly discussed.
\subsection{Extension to the Uniform Planar Array Case}
{Although the proposed Kronecker analog beamforming framework is presented assuming uniform linear array (ULA) for simplicity as in many other papers, e.g., \cite{venkateswaran2010analog,zhang2005variable,YinGesbert:CoordinatedApproachLargeScaleMIMO:2013}, we would like to point out that the extension to the case of uniform planar array (UPA) is indeed straightforward by noting that the phase response vector of a UPA can be represented as the Kronecker product of two ULA phase response vectors containing the elevation and azimuth angle-of-arrival  (see  \cite{el2014spatially,Shafin2016} for details). 
%Therefore, the proposed Kronecker decomposition can still be applied to the UPA case to perform the Kronecker-product representations to the channel and beamforming vectors, thereby enabling the Kronecker analog beamforming framework. 
The basic design principle may even be applicable to other types of arrays, e.g., circular and spherical arrays by decomposing their response vectors using appropriate basis functions, which might be Fourier series and spherical harmonics, respectively.}

\subsection{Extension to the Multi-antenna User Case}
{In this paper, the single antenna user assumption is made to simplify the exposition. Nevertheless, due to the smaller antenna spacing required in the mmWave communications, multiple antennas can also be packed at the user terminals to enable multiple data streams transmission per user. In this regard, we would like to briefly discuss the extension of the proposed hybrid beamforming scheme to the multi-antenna user case where multiple data streams per user are supported. The key idea is to integrate the proposed Kronecker analog beamforming technique and the powerful block diagonalization technique \cite{SpencerSwindleETAL:ZFsdma:2004,RajashekarTWC2016} and also the eigenmode beamforming  technique (also called SVD beamforming) to tackle the inter-cell interference, inter-user interference (within the same cell) and inter-stream interference (within the same user) in this scenario. 
Specifically, the Kronecker analog beamforming can be first designed based on the original multi-path channel to cancel the inter-cell interference and enhance the data signal paths as presented earlier.  
Next, block diagonalization can be designed according to the post-analog-processed channel to decompose the effective multiuser MIMO channel after analog beamforming to a set of parallel single-user MIMO channels, thereby eliminating the inter-user interference. 
Finally, a set of eigenmode beamformers are designed separately each targeting the corresponding effective single-user MIMO channel after block diagonalization to further divide the effective single-user MIMO into a set of orthogonal SISO sub-channels for supporting parallel multiple data streams transmission.
Note that the block diagonalization and the eigenmode beamforming are performed in the digital domain while the Kronecker analog beamforming is conducted in the analog domain. Combining all three layer processing together yields the desired hybrid beamforming.}

%the analog receive beamformer can still be designed to cancel the inter-cell interference and data path enhancement following the proposed Kronecker 

%The main steps of the proposed extension is listed below.
%
%\emph{Step1}: 
%
%It is worth pointing out that the optimal hybrid beamforming for this case is indeed not tractable due to the coupling effect between the transmit beamformer and receive beamformer as well as the non-convex unit-modulus constraints on the analog beamformer coefficients. However, a good sub-optimal solution can be developed by leveraging the block diagonalization technique [XXX]

\section{Conclusion}
This paper addressed the channel estimation and hybrid beamforming design problems for a mmWave massive MIMO system targeting the multi-cell and multiuser uplink transmission scenario, with the emphasis on tackling both the intra-cell and inter-cell interference under the hardware constraints. 
%In particular, a novel solution technique called Kronecker analog beamforming was proposed to suppress the inter-cell interference in analog domain under the uni-modulus constraints imposed on the analog beamforming coefficients due to the phase-array implementation.
%The intra-cell interference was tackled in digital domain by cascading the proposed analog beamformer with a low-dimension digital MMSE beamforming.  To provide the proposed hybrid beamforming design with required CSI, a novel channel estimation approach with low complexity was also proposed for tackling the hardware challenge called channel subspace sampling limitation and also the pilot contamination.  The proposed approach leveraged the sparsity nature of the mmWave channel and divided the channel estimation into a two-stage procedure: path AoA estimation using analog beam scanning followed by path gain estimation using analog beam steering. The decay rate of the estimation error of the proposed approach was also characterized in closed-form as a function of various key system parameters, yielding useful design guidelines for 
%addressing the pilot contamination in massive MIMO systems. 
The main contribution of this work lies in the innovative systematic design of the analog beamformer based on Kronecker decomposition, which enforces the uni-modulus constraints on the analog beamforming coefficients and allows the inter-cell interference to be cancelled in analog domain. Thanks to this, the subsequent small-scale baseband digital beamformer can focus on multiuser data streams decoupling only, enabling the minimum RF chains implementation for the hybrid beamforming design. 
Moreover, the proposed channel estimation scheme conquers the hybrid beamforming hardware limitation and achieves significant complexity reduction by customized design that exploits the channel sparsity property, allowing low-complexity implementation of the proposed hybrid beamforming design. 
Furthermore, the decay rate of the estimation error of the proposed channel estimation approach was also characterized in closed-form as a function of various key system parameters, yielding useful design guidelines for 
addressing the pilot contamination in massive MIMO systems. 
This work opens up many interesting research directions. One of them is to consider the broadband transmission system, where the resultant frequency-selective channel introduces additional challenges to inter-cell interference cancellation in analog domain and calls for dedicated design. Another interesting direction is the extension of the Kronecker analog beamforming to other antenna geometries such as circular or cylindrical arrays for developing a more general design framework.
%achieve hybrid three-dimensional (3D) beamforming.

%it is of more practical interest to co
%take into account the effect of the finite phase-shift resolution on the hybrid beamforming design and channel estimation, which calls for a more robust solution.  

\section{Acknowledgement}
The research was supported by the Hong Kong Innovation and Technology Commission under the grant GHP01213SZ, the Hong Kong Research Grants Council under the grant 17259416 and the Shenzhen-Hong Kong Innovative Technology Cooperation Funding (Grant No. SGLH20131009154139588).

\appendices
%\section{Proof of Proposition \ref{prop:snc}}\label{appendix:prop:snc}
%Note that the first constraint in problem P8 can be translated to a equivalent null-space version. Specifically, the constraint  $\text{colspan}\{{\bf F}_{\sf RF} \} \supseteq \text{colspan}\{{\bf F}_{\sf opt} \}$ is equivalent to $\text{null}\{{\bf F}_{\sf RF} \} \subseteq \text{null}\{{\bf F}_{\sf opt} \}$, where $\text{null}\{{\bf A}\}$ denote the null space of matrix ${\bf A}$. In other words, if the first constraint in P8 is satisfied, the orthogonal projection of ${\bf F}_{\sf opt}$ to the null space of  ${\bf F}_{\sf RF}$ should be equal to zero, which can be mathematically expressed as (\ref{prop:snc:1}) by noting that $({\bf I} - {\bf F}_{\sf RF}({\bf F}_{\sf RF}^H{\bf F}_{\sf RF})^{-1}{\bf F}_{\sf RF}^H)$ is exactly the orthogonal projection operation to the null space of ${\bf F}_{\sf RF}$. According to this observation, it is easy to see that the first condition in (\ref{prop:snc:1}) is to guarantee the solution within the feasible set of problem P8, and the second condition in (\ref{prop:snc:2}) is to make sure the solution ${\bf F}_{\sf RF}$ has the achievable minimum rank. Therefore, combining these two conditions together leading to the sufficient and necessary for the optimal solution for P8. To this end, the proof is completed.

\section{Proof of Lemma \ref{lemma:1}}\label{appendix:lemma:1}
According to the expression of steering vector $\bf v$ given in (\ref{sys:2}), we have
\begin{align}
\frac{|{\bf v}(\Omega)^H{\bf v}(\Phi)|}{N} &= \left|{\frac{1}{N}\sum_{m=1}^Ne^{j(m-1)(\Phi- \Omega)}}\right| =\left|{\frac{1}{N} \frac{1-e^{jN(\Phi-\Omega)}}{1-e^{j(\Phi - \Omega)}}}\right| \label{app:1} \\
& = \left|{\frac{1}{N}\frac{\sin\left({\frac{N}{2}(\Phi - \Omega)}\right)}{\sin\left({\frac{1}{2}(\Phi - \Omega)}\right)   }    }\right| \label{app:2}\\
& = \left|{\frac{\sin\left({\frac{N}{2}(\Phi - \Omega)}\right)}{\frac{N}{2}(\Phi - \Omega) } \frac{\frac{1}{2}(\Phi - \Omega)}{\sin\left({\frac{1}{2}(\Phi - \Omega)}\right)   }    }\right|\\
& = \left|{\frac{{\sf sinc}\left({\frac{N}{2}|\Phi - \Omega|}\right)}{{\sf sinc}\left({\frac{1}{2}|\Phi - \Omega|}\right)   }    }\right| \label{app:3}, 
\end{align}
%\begin{align}
%\frac{|{\bf v}(\Omega)^H{\bf v}(\Phi)|}{{\|{\bf v}(\Omega)\|^2}} &= \left|{\frac{1}{N}\sum_{m=1}^Ne^{j(m-1)(\Phi- \Omega)}}\right| =\left|{\frac{1}{N} \frac{1-e^{jN(\Phi-\Omega)}}{1-e^{j(\Phi - \Omega)}}}\right| \label{app:1} \\
%& = \left|{\frac{1}{N}\frac{\sin\left({\frac{N}{2}(\Phi - \Omega)}\right)}{\sin\left({\frac{1}{2}(\Phi - \Omega)}\right)   }    }\right| \label{app:2}\\
%& = \left|{\frac{\sin\left({\frac{N}{2}(\Phi - \Omega)}\right)}{\frac{N}{2}(\Phi - \Omega) } \frac{\frac{1}{2}(\Phi - \Omega)}{\sin\left({\frac{1}{2}(\Phi - \Omega)}\right)   }    }\right|\\
%& = \left|{\frac{{\sf sinc}\left({\frac{N}{2}|\Phi - \Omega|}\right)}{{\sf sinc}\left({\frac{1}{2}|\Phi - \Omega|}\right)   }    }\right| \label{app:3}, 
%\end{align}
where the second equality in (\ref{app:1}) is a direct result of the sum of a geometric series, (\ref{app:2}) comes from the Euler's equation that $\sin x = \frac{e^{jx} -  e^{-jx}}{2j}$, and (\ref{app:3}) results from the definition of Sinc function
${{\sf sinc} (x)} = \frac{\sin x}{x}$.

To further characterize the asymptotic behaviour of the profile of function ${\cal J}(x)$, we can start from the result in (\ref{app:3}). For $|x|\ll 1$ and $N \to \infty$, invoking the asymptotic expression of Sinc function \cite{AbramowitzBook}, it follows that, 
\begin{align} \label{app:4}
{\cal J}(x) &= \left|{ \frac{{\sf sinc}\left({\frac{N|x|}{2}}\right)}{{\sf sinc}\left({\frac{|x|}{2}}\right)   } }\right| \leq \frac{\frac{2}{N|x|}}{\left|{ 1 - O(x^2) }\right|} = \frac{2}{N|x|} \left[{ 1 + O(x^2)}\right].
%\frac{2}{N|X|} + o = O(\frac{1}{N|X|}).
\end{align}

%\begin{align} \label{app:4}
%{\cal J}(x) &= \left|{  \frac{1}{N}\frac{\sin\left({\frac{N|x|}{2}}\right)}{\sin\left({\frac{|x|}{2}}\right)   } }\right| \leq \frac{1}{\left|{ N \sin\left({\frac{|x|}{2}}\right)  }\right|} = \frac{2}{Nx} + O\left(\frac{1}{N^2}\right).
%%\frac{2}{N|X|} + o = O(\frac{1}{N|X|}).
%\end{align}
To this end, the proof is completed.

\section{Proof of Proposition \ref{Prop:PathEst:CC}}\label{appendix:Prop:PathEst:CC}
Starting from \eqref{GE_CC}, it follows that 
\begin{align}\label{app:pge:1}
|\hat a_\ell^{\sf cc}  - a_\ell| = \left| \frac{1}{N}\sum_{m\neq \ell}^La_m{\bf v}(\Phi_\ell)^H{\bf v}(\Phi_m)  +  \frac{1}{N}\sum_{n = 1}^M\beta_n{\bf v}(\Phi_\ell)^H{\bf v}(\Theta_n)  + \tilde { n}^{\sf cc}      \right|,
\end{align}
where $\tilde n^{\sf cc}$ denotes the effective noise following ${\cal CN}(0,\frac{N_0}{N})$.

Next, leveraging the result in Lemma \ref{lemma:1} and exploiting the well-known \emph{vector triangle inequality}, the expression in (\ref{app:pge:1}) can be bounded by
\begin{align}\label{app:pge:2}
|\hat a_\ell^{\sf cc}  - a_\ell| &\leq  \sum_{m\neq \ell}^L|a_m|  {\cal J}(|\Phi_\ell - \Phi_m|)  +  \sum_{n = 1}^M|\beta_n| {\cal J}(|\Phi_\ell - \Theta_n|) + O(\frac{1}{N}) \notag\\
&\leq  \frac{2}{N}\sum_{m\neq \ell}^L|a_m|  (|\Phi_\ell - \Phi_m|)^{-1} +  \frac{2}{N}\sum_{n = 1}^M|\beta_n| (|\Phi_\ell - \Theta_n|)^{-1} + O(\frac{1}{N}) \notag\\
&\leq \frac{2(L+M-1)}{N}\alpha_{\sf max}\Psi_{\sf min}^{-1}  +O(\frac{1}{N}),
\end{align}
where $\alpha_{\sf max}$ and $\Psi_{\sf min}$ follows the same definition as that in \eqref{exp:PathEst:CC}.
To this end, the proof is completed.

%where the second inequality follows the result derived in (\ref{app:4}). Note that, for a given channel realization, the parameters, $a_j,\beta_n,\Phi_\ell$ and $\Theta_n$ are constants, and the estimation error is proportional to $\frac{1}{N}$ as indicated by (\ref{app:pge:2}). Thereby, It follows that $|\hat a_\ell^{\sf LS}  - a_\ell| \leq O(\frac{1}{N})$, which complete the proof.

\section{Proof of  (\ref{esti_comp})}\label{appendix:C}
{Before deriving (\ref{esti_comp}), it is essential to first derive the analog ZF beamformer ${\bf f}_{\sf ZF}$. Given the considered simple case with $K=1$, $L=1$, $M=1$ and $N$ being even, it follows from the proposed Kronecker analog beamforming framework that the interference and data paths can be decomposed as follows
\begin{align}
{\bf v}(\Theta) &= [1, e^{j\Theta}] \otimes [1,e^{j2\Theta},e^{j4\Theta},\cdots, e^{j(N-2)\Theta}],\\
{\bf v}(\Phi) &= [1, e^{j\Phi}] \otimes [1,e^{j2\Phi},e^{j4\Phi},\cdots, e^{j(N-2)\Phi}].
\end{align}
Accordingly, the analog ZF beamformer that solves problem P7 can be written as
\begin{align}
{\bf f}_{\sf ZF} = [1, -e^{j\Theta}] \otimes [1,e^{j2\Phi},e^{j4\Phi},\cdots, e^{j(N-2)\Phi}].
\end{align}
It then follows that ${\bf f}_{\sf ZF}^H {\bf v}(\Phi) = \frac{1-e^{j(\Phi - \Theta)}}{2}N$. Note that when $|\Phi - \Theta|$ is small enough, we can easily obtain the first order approximation ${\bf f}_{\sf ZF}^H {\bf v}(\Phi) \approx \frac{-j(\Phi-\Theta)}{2}N$. As a result, the noise term in \eqref{Eq:GainEst:ZF}, i.e., $\frac{{\bf f}_{{\sf ZF}}^H{\bf n}}{{\bf f}_{{\sf ZF}}^H{\bf v}(\Phi)}$ has the variance approximately equal to $\frac{4N_0}{|\Phi - \Theta|^2N}$.

Next, following from (\ref{GE_CC}), we have
\begin{align}\label{esti_err_cc}
{\sf E} [|\hat a^{\sf cc}_\ell  - a_\ell|] &= {\sf E} \left[ \left| \beta \frac{{\bf v}(\Phi)^H{\bf v}(\Theta)}{N} + \frac{1}{\sqrt{N}}\hat n  \right| \right] \leq {\sf E} \left[ \beta \left|  \frac{{\bf v}(\Phi)^H{\bf v}(\Theta)}{N} \right| + \frac{1}{\sqrt{N}} |\hat n| \right] \notag \\
&\leq \beta \frac{2}{N|\Phi - \Theta|} + \frac{1}{\sqrt{N}} {\sf E} [|\hat n|] \leq  \beta \frac{2}{N|\Phi - \Theta|} + \frac{1}{\sqrt{N}} \sqrt{{\sf E} [\hat n^2]} \notag \\
&= \beta \frac{2}{N|\Phi - \Theta|} + \frac{\sqrt{N_0}}{\sqrt{N}},
\end{align}
where the second inequality is due to the vector triangle inequality, the third inequality is followed from Lemma \ref{lemma:1}, and the forth inequality results from the Jensen's inequality that ${\sf E}[\sqrt{x}] \leq \sqrt{{\sf E}[x]}$ for a non-negative random variable $x$.

Similarly, following from \eqref{Eq:GainEst:ZF} and invoking the Jensen's inequality we have  
\begin{align}\label{esti_err_zf}
{\sf E} [|\hat a^{\sf zf}_\ell  - a_\ell|] =  {\sf E} \left[ \left| \frac{{\bf f}_{{\sf ZF}}^H{\bf n}}{{\bf f}_{{\sf ZF}}^H{\bf v}(\Phi)}\right| \right] \leq \frac{2\sqrt{N_0}}{|\Phi - \Theta|\sqrt{N}}
\end{align}

Finally, By approximating ${\sf E} [|\hat a^{\sf cc}_\ell  - a_\ell|]$ and ${\sf E} [|\hat a^{\sf zf}_\ell  - a_\ell|]$ using their tight upper bounds derived in (\ref{esti_err_cc}) and (\ref{esti_err_zf}) respectively, the desired result presented in 
(\ref{esti_comp}) can be obtained after some simple algebra manipulation.}

\bibliographystyle{ieeetr}
\bibliography{BibDesk_File}

\begin{thebibliography}{10}

\bibitem{andrews2014will}
J.~G. Andrews, S.~Buzzi, W.~Choi, S.~V. Hanly, A.~Lozano, A.~C. Soong, and
  J.~C. Zhang, ``What will 5{G} be?,'' {\em IEEE J. Sel. Areas Commun.},
  vol.~32, pp.~1065--1082, Jun. 2014.

\bibitem{SHanCM2015}
S.~Han, C.~lin I, Z.~Xu, and C.~Rowell, ``Large-scale antenna systems with
  hybrid analog and digital beamforming for millimeter wave 5{G},'' {\em IEEE
  Commun. Mag.}, vol.~53, pp.~117--125, Apr. 2015.

\bibitem{SunTAP2013}
H.~Sun, Y.-X. Guo, and Z.~Wang, ``60-{GH}z circularly polarized u-slot patch
  antenna array on {LTCC},'' {\em IEEE Trans. Ant. Propag.}, vol.~61,
  pp.~430--435, Jan. 2013.

\bibitem{RusekLarssonMarz:ScaleUpMIMO:2012}
F.~Rusek, D.~Persson, B.~K. Lau, E.~G. Larsson, T.~L. Marzetta, O.~Edfors, and
  F.~Tufvesson, ``Scaling up {MIMO}: Opportunities and challenges with very
  large arrays,'' {\em IEEE Signal Proc. Mag.}, vol.~30, pp.~40--60, Jan. 2013.

\bibitem{GesShaETAL:FromTheoPrac:Apr:03}
D.~Gesbert, M.~Shafi, D.-S. Shiu, P.~J. Smith, and A.~Naguib, ``From theory to
  practice: An overview of {MIMO} space-time coded wireless systems,'' {\em
  IEEE J. Sel. Areas Commun.}, vol.~21, pp.~281--302, Apr. 2003.

\bibitem{Gesbert:MultiCellMIMOCooperativeNetworks:2010}
D.~Gesbert, S.~Hanly, H.~Huang, S.~Shitz, O.~Simeone, and W.~Yu, ``Multi-cell
  {MIMO} cooperative networks: A new look at interference,'' {\em IEEE J. Sel.
  Areas Commum.}, vol.~28, pp.~1380--1408, Dec. 2010.

\bibitem{SunMag2014}
S.~Sun, T.~S. Rappaport, R.~W. {Heath Jr}., A.~Nix, and S.~Rangan, ``{MIMO} for
  millimeter-wave wireless communications: Beamforming, spatial multiplexing,
  or both?,'' {\em IEEE Commun. Mag.}, vol.~52, pp.~110--121, Dec. 2014.

\bibitem{HeathCM2014}
A.~Alkhateeb, J.~Mo, N.~G. Prelcic, and R.~W. {Heath Jr.}, ``{MIMO} precoding
  and combining solutions for millimeter wave systems,'' {\em IEEE Comm. Mag.},
  vol.~52, pp.~122--131, Dec. 2014.

\bibitem{roh2014millimeter}
W.~Roh, J.-Y. Seol, J.~Park, B.~Lee, J.~Lee, Y.~Kim, J.~Cho, K.~Cheun, and
  F.~Aryanfar, ``Millimeter-wave beamforming as an enabling technology for 5{G}
  cellular communications: theoretical feasibility and prototype results,''
  {\em IEEE Commun. Mag.}, vol.~52, pp.~106--113, Feb. 2014.

\bibitem{el2014spatially}
O.~El~Ayach, S.~Rajagopal, S.~Abu-Surra, Z.~Pi, and R.~W. {Heath Jr.},
  ``Spatially sparse precoding in millimeter wave {MIMO} systems,'' {\em IEEE
  Trans. Wireless Commun.}, vol.~13, pp.~1499--1513, Mar. 2014.

\bibitem{alkhateeb2014channel}
A.~Alkhateeb, O.~El~Ayach, G.~Leus, and R.~W. {Heath Jr.}, ``Channel estimation
  and hybrid precoding for millimeter wave cellular systems,'' {\em IEEE J.
  Sel. Topics Signal Proc.}, vol.~8, pp.~831--846, Oct. 2014.

\bibitem{venkateswaran2010analog}
V.~Venkateswaran and A.-J. van~der Veen, ``Analog beamforming in {MIMO}
  communications with phase shift networks and online channel estimation,''
  {\em IEEE Trans. Signal Proc.}, vol.~58, pp.~4131--4143, Aug. 2010.

\bibitem{zhang2005variable}
X.~Zhang, A.~F. Molisch, and S.-Y. Kung, ``Variable-phase-shift-based
  {RF}-baseband codesign for {MIMO} antenna selection,'' {\em IEEE Trans.
  Signal Proc.}, vol.~53, pp.~4091--4103, Nov. 2005.

\bibitem{weiyuJSTSP2016}
F.~Sohrabi and W.~Yu, ``Hybrid digital and analog beamforming design for
  large-scale antenna arrays,'' {\em IEEE J. Sel. Topics Signal Proc.},
  vol.~10, pp.~501--513, Apr. 2016.

\bibitem{yu2016alternating}
X.~Yu, J.-C. Shen, J.~Zhang, and K.~B. Letaief, ``Alternating minimization
  algorithms for hybrid precoding in millimeter wave {MIMO} systems,'' {\em
  IEEE J. Sel. Topics Signal Proc.}, vol.~10, pp.~485--500, Apr. 2016.

\bibitem{BradyTAP2013}
J.~Brady, N.~Behdad, and A.~Sayeed, ``Beamspace {MIMO} for millimeter-wave
  communications: System architecture, modeling, analysis, and measurements,''
  {\em IEEE Trans. Ant. Propag.}, vol.~61, pp.~3814--3827, Jul. 2013.

\bibitem{WangJSAC2009}
J.~Wang, Z.~Lan, C.~Pyo, T.~Baykas, C.~Sum, M.~A. Rahman, J.~Gao, R.~Funada,
  F.~Kojima, H.~Harada, and S.~Kato, ``Beam codebook based beamforming protocol
  for multi-{G}bps millimeter-wave {WPAN} systems,'' {\em IEEE J. Sel. Areas
  Commun.}, vol.~27, pp.~1390--1399, Oct. 2009.

\bibitem{alkhateeb2015limited}
A.~Alkhateeb, G.~Leus, and R.~W. {Heath Jr.}, ``Limited feedback hybrid
  precoding for multi-user millimeter wave systems,'' {\em IEEE Trans. Wireless
  Commun.}, vol.~14, pp.~6481--6494, Nov. 2015.

\bibitem{hur2013millimeter}
S.~Hur, T.~Kim, D.~J. Love, J.~V. Krogmeier, T.~A. Thomas, and A.~Ghosh,
  ``Millimeter wave beamforming for wireless backhaul and access in small cell
  networks,'' {\em IEEE Trans. Commun.}, vol.~61, pp.~4391--4403, Oct. 2013.

\bibitem{HeathJSTSP2016}
R.~W. {Heath Jr.}, N.~Gonzalez-Prelcic, S.~Rangan, W.~Roh, and A.~M. Sayeed,
  ``An overview of signal processing techniques for millimetre wave {MIMO}
  systems,'' {\em IEEE J. Sel. Topics Signal Proc.}, vol.~10, pp.~436--453,
  Apr. 2016.

\bibitem{alkhateeb2013hybrid}
A.~Alkhateeb, O.~El~Ayach, G.~Leus, and R.~W. {Heath Jr.}, ``Hybrid precoding
  for millimeter wave cellular systems with partial channel knowledge,'' in
  {\em Information Theory and Applications Workshop (ITA), 2013}, pp.~1--5,
  IEEE, Feb. 2013.

\bibitem{RapcellularmmWave2013}
T.~S. {Rappaport et al.}, ``Millimeter wave mobile communications for 5{G}
  cellular: It will work!,'' {\em IEEE Access}, vol.~1, pp.~335--349, May 2013.

\bibitem{andrews2016modeling}
J.~G. Andrews, T.~Bai, M.~Kulkarni, A.~Alkhateeb, A.~Gupta, and R.~W. Heath~Jr,
  ``Modeling and analyzing millimeter wave cellular systems,'' {\em arXiv
  preprint arXiv:1605.04283}, 2016.

\bibitem{berraki2014application}
D.~E. Berraki, S.~M. Armour, and A.~R. Nix, ``Application of compressive
  sensing in sparse spatial channel recovery for beamforming in mmwave outdoor
  systems,'' in {\em 2014 IEEE Wireless Communications and Networking
  Conference (WCNC)}, pp.~887--892, IEEE, Apr. 2014.

\bibitem{alkhateeby2015compressed}
A.~Alkhateeby, G.~Leusz, and R.~W. {Heath Jr.}, ``Compressed sensing based
  multi-user millimeter wave systems: How many measurements are needed?,'' in
  {\em 2015 IEEE International Conference on Acoustics, Speech and Signal
  Processing (ICASSP)}, pp.~2909--2913, IEEE, Apr. 2015.

\bibitem{ramasamy2012compressive}
D.~Ramasamy, S.~Venkateswaran, and U.~Madhow, ``Compressive tracking with
  1000-element arrays: A framework for multi-{G}bps mmwave cellular
  downlinks,'' in {\em Communication, Control, and Computing (Allerton), 2012
  50th Annual Allerton Conference on}, pp.~690--697, IEEE, Oct. 2012.

\bibitem{YinGesbert:CoordinatedApproachLargeScaleMIMO:2013}
H.~Yin, D.~Gesbert, M.~Filippou, and Y.~Liu, ``A coordinated approach to
  channel estimation in large-scale multiple-antenna systems,'' {\em IEEE J.
  Sel. Areas Commun.}, vol.~31, pp.~264--273, Feb. 2013.

\bibitem{zhu2015analog}
G.~Zhu and K.~Huang, ``Analog spatial cancellation for tackling the near-far
  problem in wirelessly powered communications,'' {\em IEEE J. Sel. Areas
  Comm.}, vol.~34, pp.~1--11, Dec. 2016.

\bibitem{Randomsignalandnoise}
W.~B. {Davenport Jr.} and W.~L. Root, {\em An Introduction to the Theory of
  Random Signals and Noise}.
\newblock Wiley-IEEE Press, 1st~ed., Oct. 1987.

\bibitem{Fernandes2013}
F.~Fernandes, A.~Ashikhmin, and T.~L. Marzetta, ``Inter-cell interference in
  noncooperative {TDD} large scale antenna systems,'' {\em IEEE J. Sel. Areas
  Commun.}, vol.~31, pp.~192--201, Feb. 2013.

\bibitem{RiemannianGeometry2008}
J.~Jost, {\em Riemannian Geometry and Geometric Analysis}.
\newblock Springer, 5th~ed., Apr. 2008.

\bibitem{WeightedMMSE2011}
Q.~Shi, M.~Razaviyayn, Z.~Luo, and C.~He, ``An iteratively weighted {MMSE}
  approach to distributed sum-utility maximization for a {MIMO} interfering
  broadcast channel,'' {\em IEEE Trans. Signal Proc.}, vol.~59, pp.~4331--4330,
  Sep. 2011.

\bibitem{HorJoh:MatrAnal:85}
R.~A. Horn and C.~R. Johnson, {\em Matrix Analysis}.
\newblock London: Cambridge University Press, 1st~ed., 1990.

\bibitem{schmidt1986multiple}
R.~Schmidt, ``Multiple emitter location and signal parameter estimation,'' {\em
  IEEE Trans. Ant. Propag.}, vol.~34, pp.~276--280, Mar. 1986.

\bibitem{love2003equal}
D.~J. Love and R.~W. {Heath Jr.}, ``Equal gain transmission in multiple-input
  multiple-output wireless systems,'' {\em IEEE Trans. Commun.}, vol.~51,
  pp.~1102--1110, Jul. 2003.

\bibitem{RobustCL2016}
J.~Li, L.~Xiao, X.~Xu, and S.~Zhou, ``Robust and low complexity hybrid
  beamforming for uplink multiuser mm{W}ave {MIMO} systems,'' {\em IEEE Commun.
  Lett.}, vol.~20, pp.~1140--1143, Jun. 2016.

\bibitem{RajashekarTWC2016}
R.~Rajashekar and L.~Hanzo, ``Iterative matrix decomposition aided block
  diagonalization for mm-wave multiuser {MIMO} systems,'' {\em IEEE Trans.
  Wireless Commun.}, DOI 10.1109/TWC.2016.2628357, 2016.

\bibitem{Shafin2016}
R.~Shafin, L.~Liu, J.~Zhang, and Y.-C. Wu, ``{DoA} estimation and capacity
  analysis for {3D} millimeter wave massive-{MIMO/FD-MIMO} {OFDM} systems,''
  {\em IEEE Trans. Wireless Commun.}, vol.~15, pp.~6963--6978, Oct. 2016.

\bibitem{SpencerSwindleETAL:ZFsdma:2004}
Q.~H. Spencer, A.~L. Swindlehurst, and M.~Haardt, ``Zero-forcing methods for
  downlink spatial multiplexing in multiuser mimo channels,'' {\em IEEE Trans.
  on Signal Proc.}, vol.~52, no.~2, pp.~461 -- 471, 2004.

\bibitem{AbramowitzBook}
M.~Abramowitz and I.~A. Stegun, {\em Handbook of Mathematical Functions with
  Formulas, Graphs, and Mathematical Tables}.
\newblock New York: Dover: Dover Publications, 1965.

\end{thebibliography}

\end{document}